\documentclass[a4paper]{article}
\usepackage[margin=0.70in]{geometry}
\usepackage{amsmath,amssymb,empheq}
\usepackage{amsfonts} 
\usepackage{graphicx}
\usepackage{parskip} 
\usepackage{subcaption} 
\usepackage{multirow}	
\usepackage{float} 
\usepackage{siunitx}
\usepackage{comment}
\usepackage{todonotes}
\usepackage{mathtools}

\usepackage{multirow}

\usepackage{pgfplotstable}
\usepackage{pgfplots}

\usepackage{algorithm}
\usepackage{algorithmic}

\usepackage{xcolor}

\usepackage{authblk}

\usepackage{adjustbox}

\usepackage{url}

\DeclareMathOperator\erf{erf}

\title{Hilbert expansion based fluid models for kinetic equations describing neutral particles in the plasma edge of a fusion device}

\author[1]{V. Maes}
\author[2]{W. Dekeyser}
\author[3,4]{J. Koellermeier}
\author[2]{M. Baelmans}
\author[1]{G. Samaey}
\affil[1]{Department of Computer Science, KU Leuven, Belgium}
\affil[2]{Department of Mechanical Engineering, KU Leuven, Belgium}
\affil[3]{Bernoulli Institute for Mathematics, Computer Science and Artificial Intelligence, University of Groningen,
Nijenborgh 4, NL-9747 AG Groningen, Netherlands
}
\affil[4]{Groningen Cognitive Systems and Materials Center, University of Groningen, Nijenborgh 4, NL-9747 AG
Groningen, Netherlands
}
\date{}

\def \reponame {a \texttt{Zenodo repository}~\cite{zenodo}}

\begin{document}
\pagenumbering{arabic}
\setcounter{page}{1}
\maketitle

\section*{Abstract}
Neutral particles in the plasma edge of fusion devices based on magnetic confinement are described by a transient kinetic equation incorporating ionization, recombination, and charge-exchange collisions. In charge-exchange dominated regimes, the neutral particle velocity distribution approaches the drifting Maxwellian defined by the mean velocity and temperature of the plasma. This enables model order reduction from the kinetic equation to approximate fluid models. We derive transient fluid models consistent with the kinetic equation by exploring a splitting based approach. We split the kinetic equation in sources and sinks on the one hand, and transport combined with charge-exchange on the other hand. Combining transport with charge-exchange collisions allows for deriving Hilbert expansion based fluid models. The retrieved fluid models depend on the assumed importance (scaling) of the different terms in the split equation describing transport and charge-exchange. We explore two scalings: the hydrodynamic scaling and the diffusive scaling. The diffusive scaling fluid model closely resembles phenomenological fluid models for describing neutral particles in the plasma edge that have been derived in the past. Therefore, the Hilbert expansion based fluid models can serve as a theoretical basis for such phenomenological fluid models and elucidate some of their properties. The performance of the fluid models with respect to a discrete velocity model and a Monte Carlo reference solver is assessed in numerical experiments. The code used to perform the numerical experiments is openly available.

Keywords: plasma edge modelling, Hilbert expansion, fluid model, kinetic equation, operator splitting

\section{Introduction}
\label{sec:introduction}

Numerical modelling of the plasma edge plays an essential role in understanding and designing magnetic fusion devices~\cite{reiter_eirene_2005, kukushkin_finalizing_2011}. Plasma edge simulation codes need to account for the complex interplay between plasma transport, impurity transport,
plasma-wall interactions, and collisional processes with neutrals. The behaviour of the neutral particles, present in the plasma edge of a fusion device, is governed by a transient kinetic equation that describes particle transport and several collisional processes~\cite{reiter_eirene_2005, horsten_fluid_2019, mortier_advanced_2020}. The solution of this kinetic equation is called the particle velocity distribution and determines how the neutral particles are distributed in space and velocity throughout time.

Modelling the neutral particles poses two main difficulties. First, the kinetic description of the neutral particle behaviour is high-dimensional, as both the position and the velocity of the particles are resolved. Second, the rapid succession of particle interactions in high-collisional regimes results in stiff dynamics~\cite{klar_asymptotic-induced_1998}. To reduce the computational cost when simulating the neutral particle behaviour, model order reduction techniques can be considered that avoid resolving the particle velocities. These reduced models are called fluid models. Ideally, a fluid model is equipped with an approximation of the particle velocity distribution, from which a variety of macroscopic quantities of interest (QoIs) can be estimated. These QoIs, e.g., the particle, momentum, and energy density of the neutral particles, can then be expressed as integrals of this approximate particle velocity distribution over velocity space. 

The literature on fluid models describing neutrals in the plasma edge mainly focuses on solving steady-state problems using models based on a variant of the Method of Moments~\cite{horsten_comparison_2016, van_uytven_implementation_2020, van_uytven_assessment_2022, vold_neutral_1990, vold_coupled_1992,taroni_multi-fluid_1992, valentinuzzi_fluid_2018, prinja_first_1992, rognlien_fully_1992, rognlien_2-d_1994, wising_simulation_1996, wising_simulation_1996-1, rensink_comparison_1998, umansky_modeling_2003}. These fluid models typically introduce approximations and simplifications (e.g., discarding terms or adding additional constraints) in the fluid description based on an intuitive, phenomenological basis. Incorporating knowledge of domain experts, these models are accurate given the right physical conditions, but have a rather unclear range of validity and offer no consistent approximation to the particle velocity distribution. Such phenomenological fluid models have been developed in 4 of the major tokamak plasma edge modelling code bases: B2(.5)-EIRENE~\cite{horsten_comparison_2016, van_uytven_implementation_2020, van_uytven_assessment_2022}, SolEdge2D-EIRENE~\cite{bufferand_numerical_2015, valentinuzzi_fluid_2018}, EDGE2D-NIMBUS~\cite{vold_neutral_1990, vold_coupled_1992,taroni_multi-fluid_1992}, and UEDGE-DEGAS~\cite{prinja_first_1992, rognlien_fully_1992, rognlien_2-d_1994, wising_simulation_1996, wising_simulation_1996-1, rensink_comparison_1998}, where they provide an alternative to the more involved  Monte Carlo codes (EIRENE, NIMBUS, and DEGAS) for treating the neutral particles. Similar fluid models have also been implemented in codes for stellarator plasma edge modelling, such as the BoRiS code~\cite{umansky_modeling_2003}. In this paper, we attempt to obtain fluid models that are similar to the phenomenological fluid models that have been derived in the past, but based on a more systematic, mathematically rigorous approach. As we will see, following a systematic approach clarifies some of the properties of the fluid models, such as their range of validity and their approximate particle velocity distribution.

The difficulties introduced by high dimensionality and high collisionality are not restricted to the setting of neutrals in the plasma edge, but arise in many applications such as the kinetic modelling of rarefied gases~\cite{saint-raymond_hydrodynamic_2009}, chemotaxis~\cite{othmer_diffusion_2000, othmer_diffusion_2002}, neutron transport, and radiative transfer~\cite{pomraning_diffusive_1992}. Consequently, there exists a vast literature on the derivation of fluid models from kinetic equations. Two classical approaches lead to fluid models equipped with an approximate particle velocity distribution. The first approach, the Method of Moments, as pioneered by Grad~\cite{grad_kinetic_1949, flugge_principles_1958}, selects a function space based on an ansatz and approximates the particle velocity distribution by an infinite expansion in the basis functions of that space. A moment model is then defined by a suitable closure (e.g., truncating the expansion after a finite number of terms) that allows for the derivation of a closed set of fluid equations for the expansion coefficients (the so-called moments). The method is rooted in approximation theory and has some interesting properties~\cite{fan_model_2016, struchtrup_macroscopic_2005} if the moment model converges to the solution of the kinetic equation for increasing number of moments. There is, however, no guarantee that the moment model will converge~\cite{cai_holway-weiss_2019}. One of the main advantages of a convergent moment model is that the performance can easily be improved by adding more terms to the closed expansion. A drawback of the method is that the chosen function space and closure determine the assumptions that are included in the moment model, but the physical interpretation of these assumptions is not always directly clear. A second drawback is that the stiffness of the dynamics in high-collisional regimes has to be solved by specialized techniques, such as projective integration~\cite{koellermeier_projective_2021, cusicanqui_subm_2022}.

The second classical approach explores the behaviour of the system close to equilibrium by using perturbation theory~\cite{Van_Dyke_Perturbation_1975}. The approach consists of deriving a so-called Hilbert expansion around the equilibrium state that leads to a closed set of fluid equations~\cite{saint-raymond_hydrodynamic_2009, pomraning_diffusive_1992}. A Hilbert expansion is derived by scaling each of the terms in the kinetic equation using a non-dimensional scaling parameter $\varepsilon \ll 1$ to a given power. The scaling parameter appears after suitable non-dimensionalisation of the kinetic equation and explicitly and systematically introduces physical assumptions about which terms are large and which are small. The particle velocity distribution itself is also expanded in terms of $\varepsilon$. The idea is that in the limit $\varepsilon \rightarrow 0$, terms with a given order in $\varepsilon$ have to balance each other as terms with another order in $\varepsilon$ are an order of magnitude larger or smaller and therefore can never balance the terms under consideration, i.e., terms with different orders in $\varepsilon$ decouple. This decoupling constitutes a mathematically rigorous way to derive fluid models close to equilibrium, where taking the limit $\varepsilon \rightarrow 0$ can be interpreted as an idealization of the introduced physical assumptions. The dynamics of the resulting fluid models is independent of $\epsilon$ and does not become stiff in the high-collisional limit.

Applying model order reduction techniques, such as the Method of Moments or a Hilbert expansion based approach, directly to complicated kinetic equations can lead to intricate fluid models with a narrow range of validity. One way around this problem is to introduce operator splitting~\cite{glowinski_operator_2016}. Operator splitting allows to split kinetic equations in several parts which can then be treated separately. One such approach splits the kinetic equation in transport and event equations, where events can comprise collisions, sources, and sinks~\cite{dimarco_hybrid_2006, dimarco_asymptotic-preserving_2018}. The split equations can then be treated separately, e.g., by a moment model. The event equations typically pose no difficulties, as the different moments are independent of each other, leading to a closed set of fluid equations. The moments of the transport equation, however, are coupled in such a way that an infinite chain of fluid equations is generated, which has to be solved by introducing a closure.

The goal of this paper is to derive transient fluid models, equipped with an approximate particle velocity distribution and a clear range of validity, that accurately describe the neutral particle behaviour in the plasma edge in high-collisional regimes and can be used as a rigorous basis for the development of hybrid fluid/Monte Carlo methods~\cite{borodin_fluid_2021, valentinuzzi_two-phases_2019, dimarco_hybrid_2008, karney_modeling_1998}. The followed approach splits the kinetic equation in transport combined with charge-exchange collisions on the one hand, and sources and sinks on the other hand. Instead of taking moments of the equation describing transport and charge-exchange collisions, our focus will be on a Hilbert expansion based route towards fluid models. The Hilbert expansions treat introduced assumptions in a clear and rigorous way, allowing for statements about accuracy and range of validity. Additionally, the Hilbert expansion has an explicit expression for its particle velocity distribution approximation underlying the resulting fluid model. This approximate particle velocity distribution can then be used to evaluate a variety of velocity dependent QoIs, without actually having to resolve the velocity of the neutral particles as is done when solving the kinetic equation directly. The fluid models are derived in the transient setting, allowing for transient simulations of the plasma edge. The focus of this paper is on the bulk of the plasma edge, i.e., without the typical, complex boundary interactions~\cite{horsten_comparison_2016, horsten_fluid_2019} describing amongst others the interactions with the divertor targets. We therefore only implement periodic boundary conditions and leave the more advanced boundary conditions for future work.

The rest of the paper is structured as follows. Section~\ref{sec:bulk} describes the kinetic equation governing the neutral particle behaviour in more detail and explains the splitting approach used for deriving the fluid models. Section~\ref{sec:HE_Fluidmodels} treats the Hilbert expansion based fluid models for the split equation describing transport and charge-exchange. Next, in Section~\ref{sec: solution_procedure}, we discuss how the different fluid models and estimators for the QoIs can be discretized and implemented. Verification of the derived fluid models is done by comparison with a discrete velocity model~\cite{mieussens_discrete_2000} and a particle tracing Monte Carlo method~\cite{mortier_advanced_2020, spanier_monte_1969}. In Section~\ref{sec: experiments}, three numerical experiments are performed to showcase the performance of the fluid models compared to the two reference solutions. Finally, Section~\ref{sec: conclusion} summarizes the conclusions of this paper. The code accompanying this paper is openly available in \reponame.

\section{Kinetic equation describing neutrals in the plasma edge}
\label{sec:bulk}
This paper focusses on a simplified transient kinetic equation describing neutral particles in a plasma background medium with a 1D physical and velocity space: $x \in D \subset \mathbb{R}$, $v \in \mathbb{R}$, $t \geq 0$. The domain $D$ in which we want to solve the kinetic equation can be subdivided in the bulk and the boundary, where the boundary treatment, in general, can be quite complex~\cite{horsten_comparison_2016, horsten_fluid_2019}. 
In this paper, we focus on the neutral particle description in the bulk of the domain by taking periodic boundary conditions. The kinetic equation for which we derive fluid models reads as follows:
\begin{equation}
\begin{split}
&\underbrace{\partial_t f(x,v,t)}_{\substack{\text{transient}}} + \underbrace{v \partial_x f(x,v,t)}_{\substack{\text{transport}}} = \underbrace{S(x,v,t)}_{\substack{\text{source}}} - \underbrace{R_i(x,t) f(x,v,t)}_{\substack{\text{ionization sink}}} \\ 
& \hspace{0.5em} +\underbrace{R_{cx}(x,t) \left( M(v \mid x,t) \int f(x,v',t)dv' - f(x,v,t) \right)}_{\substack{\text{charge-exchange collision operator}}},
\end{split}
\label{eq:kinetic_equation}
\end{equation}
where $f(x,v,t)$, the particle velocity distribution, represents the density of particles at time $t$ with a given position $x$ and velocity $v$; $S(x,v,t)$ is the source describing how many neutrals are created at a position $x$ with velocity $v$ per unit of time; the ionization rate $R_i(x,t)$ and charge-exchange rate $R_{cx}(x,t)$, are known functions that depend on the background plasma state~\cite{horsten_comparison_2016}; $M(v \mid x,t)$ is the normalized drifting Maxwellian describing the velocity distribution of the ions in the background plasma with mean velocity $u_p(x,t)$ and variance $\sigma_p^2(x,t)$ (related to the ion temperature $T_i(x,t)$):
\begin{equation}
M(v \mid x,t) = \frac{1}{\sqrt{2 \pi \sigma_p^2(x,t)}} \exp \left(-\frac{1}{2}\frac{(v-u_p(x,t))^2}{\sigma_p^2(x,t)} \right).
\label{eq:Maxwellian}
\end{equation}
This Maxwellian represents the post-collisional velocity distribution of the neutral particles: after a charge-exchange collision between a neutral and an ion plasma particle, the neutral gets a velocity that is distributed according to the plasma velocity distribution~\cite{mortier_advanced_2020}.
The moments $u_p(x,t)$ and $\sigma_p^2(x,t)$ of the Maxwellian are known from the plasma, rendering the kinetic equation~\eqref{eq:kinetic_equation} linear. In the literature, linear kinetic equations are sometimes referred to as linear transport equations~\cite{papanicolaou_asymptotic_1975, bensoussan_boundary_1979, lapeyre_introduction_2003, othmer_diffusion_2000, othmer_diffusion_2002, case_linear_1967}.

In neutral particle modelling, we are interested in some QoIs that are typically moments of the particle velocity distribution $f(x,v,t)$, i.e., integrals over velocity space of the following form:
\begin{equation}
Q(x,t) = \int q(v)f(x,v,t)dv,
\label{eq:QoI_Generic}
\end{equation} 
where $q(v)$ is a function of velocity. In this paper, we focus on the neutral particle density $\rho(x,t)$, momentum density $m(x,t)$, and energy density $E(x,t)$, the three lowest order moments of the particle velocity distribution, which are defined as follows:
\begin{equation}
\rho(x,t) = \int f(x,v,t)dv,\quad m(x,t) = \int v f(x,v,t)dv,\quad 
E(x,t) = \int \frac{v^2}{2}f(x,v,t)dv,
\label{eq:QoIs}
\end{equation}
where $m(x,t) := \rho(x,t) u(x,t)$ and $2E(x,t) := \rho(x,t) (u(x,t)^2 + \sigma^2(x,t))$. In these relations, $u(x,t)$ and $\sigma^2(x,t)$ denote the mean velocity and variance on the velocity of the neutral particles. In the notation of equation~\eqref{eq:QoI_Generic}, we have $Q(x,t) = \{\rho(x,t), m(x,t), E(x,t)\}$ and $q(v) = \{1,v,v^2/2\}$.

\subsection{Splitting of the kinetic equation}
To obtain a Hilbert expansion based fluid model, the relative importance of the effects included in the kinetic equation~\eqref{eq:kinetic_equation} has to be assessed. As will be discussed in Section~\ref{sec:HilbertExpansions}, deciding on the importance of the different terms in a complex kinetic equation tends to be a convoluted endeavour. To facilitate the derivation of a fluid model, first the kinetic equation is split in sources~\eqref{eq:source_equation}, conservative processes~\eqref{eq:conservation_equation} (transport and charge-exchange in this case), and sinks~\eqref{eq:sink_equation} (ionization in this case):
\begin{subequations}\label{eq:system}
\begin{empheq}[left=\empheqlbrace]{align}
  &\partial_t f(x,v,t) = S(x,v,t) \label{eq:source_equation}
  \\
  &\partial_t f(x,v,t) + v\partial_x f(x,v,t) =
 R_{cx}(x,t)\left( M(v\mid x,t) \int f(x,v',t)dv' - f(x,v,t) \right) \label{eq:conservation_equation}
  \\
  &\partial_t f(x,v,t) = -R_i(x,t) f(x,v,t). \label{eq:sink_equation}
\end{empheq}
\end{subequations}

The performed operator splitting introduces a first order discretization error in time~\cite{glowinski_operator_2016, lapeyre_introduction_2003}. Note that such a time discretization error will be present in the discretized fluid models anyway, so this is not problematic. Let us first discuss the source equation~\eqref{eq:source_equation} and ionization equation~\eqref{eq:sink_equation}. These equations only contain a derivative with respect to time and can therefore be solved using a simple time stepping scheme. Discretizing with forward Euler, the approximate solution of the source equation~\eqref{eq:source_equation} after a time step $\Delta t$ can be written as:
\begin{equation}
f(x,v,t_0+\Delta t) \approx f(x,v,t_0) + S(x,v,t_0) \Delta t.
\end{equation}
The approximate solution of the ionization equation~\eqref{eq:sink_equation} after a time step $\Delta t$ can be written as:
\begin{equation}
    f(x,v,t_0 + \Delta t) \approx \exp(-R_i(x,t_0) \Delta t) f(x,v,t_0).
\end{equation}
The interpretation of this equation is simply that a fraction of particles $\exp(-R_i(x,t_0) \Delta t)$ did not undergo an ionization collision during the time $\Delta t$ and is still present in the domain at time $t_0 + \Delta t$. A fraction $( 1 - \exp(-R_i(x,t_0) \Delta t) )$ of the particles did undergo an ionization collision during the time $\Delta t$ and disappeared in the ionization sink.

Obtaining a fluid model version of the source and ionization equation requires taking moments of the split kinetic equations. By taking moments of these equations, we retrieve the split fluid equations governing the influence of the source and ionization on the macroscopic QoIs. For the source equation~\eqref{eq:source_equation} we find:
\begin{equation}
        Q(x,t_0 + \Delta t) \approx Q(x,t_0) + \Delta t \int q(v)S(x,v,t_0) dv.
 \label{eq:moments_of_source}
\end{equation}
When the velocity dependence of the source $S(x,v,t)$ is known, e.g., a local drifting Maxwellian~\eqref{eq:Maxwellian}, the integral on the right hand side can readily be evaluated. For the ionization equation~\eqref{eq:sink_equation} we find:
\begin{equation}
        Q(x,t_0 + \Delta t) \approx \exp(-R_i(x,t_0) \Delta t) Q(x,t_0),
    \label{eq:moments_of_i}
\end{equation}
showing that a fraction of the QoIs is lost in the ionization sink.

The main difficulties in obtaining a fluid model from the split equations~\eqref{eq:source_equation}-\eqref{eq:sink_equation} are posed by the conservation equation~\eqref{eq:conservation_equation}. We treat this equation using a Hilbert expansion in the next section.

\section{Hilbert expansion based fluid models}
\label{sec:HE_Fluidmodels}
In high-collisional regimes, the charge-exchange collision operator on the right hand side of the conservation equation~\eqref{eq:conservation_equation} defines a collisional equilibrium, which enables model order reduction of the kinetic equation to a fluid model by means of a Hilbert expansion. Note that the charge-exchange collision operator is a linear BGK-like collision operator~\cite{bhatnagar_model_1954} describing neutral particles that only collide with the background plasma and not with themselves. The mean velocity and variance of the Maxwellian~\eqref{eq:Maxwellian} are therefore defined by the plasma ($u_p(x,t)$, $\sigma^2_p(x,t)$). This is in contrast to the Boltzmann-BGK collision operator for self-collisions in rarefied gas dynamics, where the mean velocity and variance of the Maxwellian correspond to the ones of the particle velocity distribution $f(x,v,t)$ itself, making the collision operator non-linear~\cite{saint-raymond_hydrodynamic_2009}. Additionally, the Boltzmann-BGK collision operator has three collisional invariants (conservation of particles, momentum, and energy), whereas the charge-exchange collision operator only has one (conservation of particles), as momentum and energy are exchanged with the background plasma (see Appendix~\ref{sec: app: momene exchange}). These properties of the charge-exchange collision operator lead to different fluid models than the typical Euler or Navier-Stokes equations obtained for Boltzmann equations describing rarefied gases with self-collisions~\cite{saint-raymond_hydrodynamic_2009, pomraning_diffusive_1992}.

\subsection{Hilbert expansions}
\label{sec:HilbertExpansions}
Hilbert expansions have three main assumptions:
\begin{itemize}
\item Some terms in a kinetic equation are more important than others. 
\item Derivatives are sufficiently mild.
\item The particle velocity distribution $f(x,v,t)$ is close to an equilibrium. 
\end{itemize}
To make these assumptions mathematically rigorous, a small scaling parameter $0 \leq \varepsilon \ll 1$ is introduced. Each term of the kinetic equation is assigned a scaling $\varepsilon^k$ with $k \in \mathbb{Z}$, where increasing the exponent corresponds to decreasing the importance of that term. The physical interpretation is that we normalize each quantity $g$ as $g = \varepsilon^k \tilde{g}$ such that $\tilde{g} \sim \mathcal{O}(1)$, meaning that the magnitude of each quantity is captured explicitly by the scaling $\varepsilon^k$. The assumption that derivatives are sufficiently mild enforces that they do not change a quantity's order in $\varepsilon$, e.g., $\partial_x (\varepsilon^k g(x,v,t)) = \varepsilon^k \partial_x g(x,v,t) $ (if a function $g(x,v,t)$ is of order $k$, then its derivatives are as well). A set of physical assumptions on the importance of different terms in a kinetic equation is called a scaling. Two such scalings are treated in Section~\ref{subsec:Diffusive_scaling} and~\ref{subsec:Hydrodynamic_scaling}. Typically, kinetic equations are assumed to be dominated by the collision operator~\cite{papanicolaou_asymptotic_1975, othmer_diffusion_2000, saint-raymond_hydrodynamic_2009, pomraning_diffusive_1992, bensoussan_boundary_1979}, resulting in a collision dominated equilibrium around which an expansion can be constructed. We define a Hilbert expansion: 
\begin{equation}
f(x,v,t) \approx f_0(x,v,t) + \varepsilon f_1(x,v,t) + \varepsilon^2 f_2(x,v,t) + \hdots \quad,
\label{eq:Hilbert_expansion_definition}
\end{equation} 
where $f_0(x,v,t)$ represents the equilibrium and $f_i(x,v,t)$, $i \geq 1$, higher order perturbations. We then insert this Hilbert expansion in the scaled kinetic equation. Taking the limit $\varepsilon \rightarrow 0$ where $\varepsilon^k \gg \varepsilon^{k+1}$, terms with a different order $k$ in $\varepsilon$ decouple as they cannot influence each other. This limit, a mathematical idealization of the made assumptions, leads in a natural way to a closed set of macroscopic evolution equations: a fluid model.

Closed fluid models have as many evolution equations as degrees of freedom. The only degrees of freedom in a Hilbert expansion are those present in the collision dominated equilibrium $f_0(x,v,t)$. The Boltzmann-BGK collision operator has three degrees of freedom: $\rho(x,t)$, $u(x,t)$, and $\sigma^2(x,t)$, so in that case fluid models consist of three evolution equations. The charge-exchange collision operator only has one degree of freedom: $\rho(x,t)$, so the resulting fluid models consist of just one evolution equation. We can therefore reduce the high-dimensional linear kinetic equation to a single macroscopic evolution equation. Other velocity dependent QoIs follow from inserting the Hilbert expansion ansatz~\eqref{eq:Hilbert_expansion_definition}  in the velocity space integral~\eqref{eq:QoI_Generic}. 

Only having transport and charge-exchange in~\eqref{eq:conservation_equation} allows for making clear statements about the importance of the different terms by means of the scaling parameter $\varepsilon$. Putting more effects in the Hilbert expansion requires the introduction of more and more assumptions on the relative importance of the terms, narrowing down the range of validity of the fluid model, as the extent to which the assumptions hold decreases. That is why we split the source and sink terms, for which taking moments does not result in an infinite hierarchy of equations, from the kinetic equation that is reduced using a Hilbert expansion (see Section~\ref{sec:bulk}).

\subsection{Fredholm alternative and self-adjoint operators}
During the derivation of Hilbert expansion based fluid models, so-called solvability conditions are encountered. To construct these solvability conditions, the Fredholm alternative can be exploited. The Fredholm alternative states that for self-adjoint operators $\mathcal{L}$, for which the null space is orthogonal to the range, an inhomogeneous equation $\mathcal{L}f = g$ only has a solution $f$ ($g$ lies in the range of $\mathcal{L}$) if $\langle g, h \rangle = 0$ for all $h$ in the null space of the operator~\cite{hutson_applications_1980, maes_subm_2022}. To apply this to the Hilbert expansion derivations, we first search for a weighted inner product for which the charge-exchange collision operator
\begin{equation}
\mathcal{L}f(v) \coloneqq R_{cx} \left( M(v)\int f(v')dv' - f(v) \right)
\label{eq:linear_BGK_type_operator}
\end{equation} 
is self-adjoint (we ignore the dependence on $x,t$ for a moment as the collision operator only acts on $v$). The required weighted inner product, which we denote by $\langle \cdot,\cdot \rangle_\mathcal{L}$, has $w(v) = M(v)^{-1}$ as weight function:
\begin{equation}
\langle f(v), g(v) \rangle_\mathcal{L} \coloneqq \int f(v)g(v)M(v)^{-1}dv.
\end{equation} 
For that choice of weight function, we indeed obtain that~\eqref{eq:linear_BGK_type_operator} is self-adjoint:
\begin{equation}
\langle \mathcal{L}f(v), g(v) \rangle_\mathcal{L}  = \langle f(v), \mathcal{L}g(v) \rangle_\mathcal{L}.
\end{equation}

Using this inner product, the Fredholm alternative can be used to determine under which conditions inhomogeneous equations of the form $\mathcal{L}f(v) = g(v)$ are solvable. Given the functions $h(v)$ that span the null space of the self-adjoint operator: $\mathcal{L}h(v) = 0$, the Fredholm alternative dictates that $\langle g(v),h(v) \rangle_\mathcal{L} = 0$ has to hold, for each conceivable $h(v)$, for $g(v)$ to be in the range of the operator $\mathcal{L}$. These inner products that have to be zero are constraints that we call solvability conditions.

For the charge-exchange collision operator~\eqref{eq:linear_BGK_type_operator}, the null space (for $R_{cx} \neq 0$) is governed by the following equation:
\begin{equation}
\begin{split}
0 &= R_{cx} \left(M(v) \int h(v')dv' - h(v) \right)\\
\Rightarrow \quad h(v) &= M(v) \int h(v') dv'.
\end{split}
\label{eq:nullspace_cxoperator}
\end{equation}
Without loss of generality, normalizing the functions spanning the null space
\begin{equation}
\int h(v) dv = 1
\end{equation}
results in $h(v) = M(v)$.

It follows that the solvability condition for functions $g(v)$ to be in the range of the charge-exchange collision operator is given by:
\begin{equation}
\langle g(v), h(v) \rangle_\mathcal{L} = \int g(v)M(v)M(v)^{-1}dv = \int g(v)dv = 0.
\label{eq:solvability_BGK-likeCollisionOperator}
\end{equation}
The solvability condition dictates that only zero mean functions are in the range of the charge-exchange collision operator.

\subsection{Diffusive scaling Hilbert expansion}
\label{subsec:Diffusive_scaling}
We now turn to the derivation of a Hilbert expansion based fluid model for the split conservation equation~\eqref{eq:conservation_equation} in the so-called diffusive scaling. A scaling refers to a set of physical assumptions that determines with which order in $\varepsilon$ the different terms in the kinetic equation are scaled. The physical assumptions in the diffusive scaling are as follows: 
\begin{itemize}
\item The plasma particle velocities are high (see Appendix~\ref{sec: app: velocity scaling}): $v_p \sim \mathcal{O}(1/\varepsilon)$, but the mean plasma particle velocity is relatively low: $u_p(x,t) \sim \mathcal{O}(1)$. As a result, the peculiar velocities are high: $c_p(x,t) = v_p - u_p(x,t) \sim \mathcal{O}(1/\varepsilon)$, meaning that the variance on the plasma particle velocities scales as: $\sigma_p^2(x,t) = \int c_p(x,t)^2 M(v \mid x,t)dv \sim \mathcal{O}(1/\varepsilon^2)$. We are thus describing a relatively slow plasma with a high temperature. Assuming that the neutral particles are almost in equilibrium with the plasma particles such that the neutral particle velocities $v$ have the same order of magnitude as the plasma particle velocities $v_p$, we introduce the following scaled variables: $\tilde{v}=\varepsilon v$, $\tilde{\sigma}_p^2(x,t) = \varepsilon^2 \sigma_p^2(x,t)$.

\item Even though the neutral particle velocities are high, the mean free path~\cite{saint-raymond_hydrodynamic_2009} $\lambda$ has to be small, because otherwise the equation would not be collision dominated and a fluid model would not be sensible. Small values for $\lambda$ are achieved by assuming large values for the collision rate $R_{cx}(x,t)$. The requirement that $\lambda \sim \mathcal{O}(\varepsilon)$, such that $\lambda \rightarrow 0$ as $\varepsilon \rightarrow 0$, leads to the following scaling for the collision rate: $\lambda \sim v/R_{cx}(x,t) \sim \mathcal{O}(\varepsilon) \Leftrightarrow R_{cx}(x,t) \sim \mathcal{O}(1/\varepsilon^2)$. We introduce the scaled variable: $\tilde{R}_{cx}(x,t) = \varepsilon^2 R_{cx}(x,t)$.
\end{itemize}

\textbf{Remark 1:} The scaled Maxwellian reads as follows:
\begin{equation}
\tilde{M}(\tilde{v} \mid x,t) = \frac{\varepsilon}{\sqrt{2 \pi \tilde{\sigma}_p^2(x,t)}} \exp \left(-\frac{1}{2}\frac{(\tilde{v}-\varepsilon u_p(x,t))^2}{\tilde{\sigma}_p^2(x,t)} \right) 
= \varepsilon M(\tilde{v} \mid x,t) .
\label{eq:scaled_Maxwellian}
\end{equation}
This can easily be verified by inserting the definitions of $\tilde{v}$ and $\tilde{\sigma}_p^2(x,t)$ into the expression. The scaled Maxwellian thus has mean $\varepsilon u_p(x,t)$ and variance $\tilde{\sigma}_p^2(x,t)$. Note that the Maxwellian is scaled by $\varepsilon$ due to the presence of the scaled variance in the normalization constant.

Inserting the diffusive scaling into the kinetic conservation equation~\eqref{eq:conservation_equation} and multiplying by $\varepsilon^2$ results in:
\begin{equation}
\varepsilon^2 \partial_t f(x,\tilde{v},t) + \varepsilon \tilde{v} \partial_x f(x,\tilde{v},t) = \tilde{R}_{cx}(x,t) \left( M(\tilde{v} \mid x,t)\int f(x,\tilde{v}',t)d\tilde{v}' - f(x,\tilde{v},t) \right).
\label{eq:diffusive_scaled_eq_v}
\end{equation}
Recall that the mean plasma velocity $u_p(x,t)$ scales differently than the peculiar velocity $c_p(x,t)$. Therefore, we decompose the transport term in a mean part and centered (zero-mean) part with respect to the scaled Maxwellian~\eqref{eq:scaled_Maxwellian} as explained in Refs.~\cite{papanicolaou_asymptotic_1975, lapeyre_introduction_2003}: 
\begin{equation}
\tilde{v} = \varepsilon u_p(x,t) + \tilde{c}(x,t),
\label{eq:velocity_decomp}
\end{equation}
such that all scalings are explicitly present in the equation. Inserting this decomposition in~\eqref{eq:diffusive_scaled_eq_v} yields:
\begin{equation}
\begin{split}
&\varepsilon^2 \partial_t f(x,\tilde{v},t) + \varepsilon^2 \partial_x \left( u_p(x,t)f(x,\tilde{v},t) \right) + \varepsilon  \partial_x \left(\tilde{c}(x,t) f(x,\tilde{v},t)\right) \\
& \hspace{0.5em} = \tilde{R}_{cx}(x,t) \left( M(\tilde{v} \mid x,t)\int f(x,\tilde{v}',t)d\tilde{v}' - f(x,\tilde{v},t) \right).
\label{eq:scaled_kinetic_equation}
\end{split}
\end{equation}
Because of the high collisionality, the neutral particles will be close to their local equilibrium state. Therefore, we propose a Hilbert expansion~\eqref{eq:Hilbert_expansion_definition} for the particle velocity distribution. Inserting the Hilbert expansion in the scaled kinetic equation~\eqref{eq:scaled_kinetic_equation} and writing equations per order in $\varepsilon$ results in:
\begin{equation}
\begin{split}
\varepsilon^0:& \hspace{0.5em} 0 = \tilde{R}_{cx}(x,t) \left( M(\tilde{v} \mid x,t)\int f_0(x,\tilde{v}',t)d\tilde{v}' - f_0(x,\tilde{v},t) \right)\\
\varepsilon^1:& \hspace{0.5em}  \partial_x \left(\tilde{c}(x,t) f_0(x,\tilde{v},t)\right) = \tilde{R}_{cx}(x,t) \left( M(\tilde{v} \mid x,t)\int f_1(x,\tilde{v}',t)d\tilde{v}' - f_1(x,\tilde{v},t) \right)\\
\varepsilon^2:& \hspace{0.5em} \partial_t f_0(x,\tilde{v},t) + \partial_x \left( u_p(x,t)f_0(x,\tilde{v},t) \right) + \partial_x \left(\tilde{c}(x,t) f_1(x,\tilde{v},t)\right) \\
& \quad = \tilde{R}_{cx}(x,t) \left( M(\tilde{v} \mid x,t)\int f_2(x,\tilde{v}',t)d\tilde{v}' - f_2(x,\tilde{v},t) \right)\\
\vdots\\
\varepsilon^k:& \hspace{0.5em} \partial_t f_{k-2}(x,\tilde{v},t) + \partial_x \left( u_p(x,t)f_{k-2}(x,\tilde{v},t) \right) + \partial_x \left(\tilde{c}(x,t) f_{k-1}(x,\tilde{v},t)\right)\\
& \quad = \tilde{R}_{cx}(x,t) \left( M(\tilde{v} \mid x,t)\int f_k(x,\tilde{v}',t)d\tilde{v}' - f_k(x,\tilde{v},t) \right).
\end{split}
\end{equation}
In the limit for $\varepsilon \rightarrow 0$, these equations decouple. The $\varepsilon^0$-equation is called the leading order equation and only contains the equilibrium $f_0(x,\tilde{v},t)$. Solving this equation therefore determines the equilibrium. Note that the leading order equation is the same as equation~\eqref{eq:nullspace_cxoperator}, meaning that the equilibrium resides in the null space of the charge-exchange collision operator. Defining the particle density of the equilibrium function as $\rho_0(x,t) \coloneqq \int f_0(x,\tilde{v},t) d\tilde{v}$, the equilibrium follows from a straightforward manipulation:
\begin{equation}
f_0(x,\tilde{v},t) = \rho_0(x,t)M(\tilde{v} \mid x,t).
\label{eq:f0_Diffusive}
\end{equation}

To obtain the first order perturbation, we insert the result for $f_0(x,\tilde{v},t)$ in the $\varepsilon^1$-equation (first order equation), resulting in:
\begin{equation}
\partial_x \left(\tilde{c}(x,t) \rho_0(x,t) M(\tilde{v} \mid x,t) \right) = \tilde{R}_{cx}(x,t) \left( M(\tilde{v} \mid x,t)\int f_1(x,\tilde{v}',t)d\tilde{v}' - f_1(x,\tilde{v},t) \right).
\end{equation}
This is an inhomogeneous integral equation. To have a solution $f_1(x,\tilde{v},t)$, the left hand side of the equation has to be in the range of the charge-exchange collision operator on the right hand side. The solvability condition is derived above in equation~\eqref{eq:solvability_BGK-likeCollisionOperator}. Inserting the left hand side into the solvability condition results in the following constraint:
\begin{equation}
\begin{split}
\int \partial_x \left(\tilde{c}(x,t) \rho_0(x,t) M(\tilde{v} \mid x,t) \right) d\tilde{v} &= 0\\
\partial_x \left( \rho_0(x,t) \int (\tilde{v}-\varepsilon u_p(x,t)) M(\tilde{v} \mid x,t) d\tilde{v} \right) &= 0\\
\partial_x \left( \rho_0(x,t) \left( \varepsilon u_p(x,t) - \varepsilon u_p(x,t) \right) \right) &= 0\\
0 &= 0. \quad \checkmark
\label{eq:firstorder_solvability_diff_scaling}
\end{split}
\end{equation}

\textbf{Remark 2:} By writing out the scaled velocity $\tilde{v}$ as the sum of the mean part $\varepsilon u_p(x,t)$ and peculiar part $\tilde{c}(x,t)$ in~\eqref{eq:velocity_decomp}, the solvability condition is fulfilled automatically. If we had not done this, the ‘hidden' scaling of the two components of the velocity would have led us to the condition $\partial_x(\rho_0(x,t) u_p(x,t)) = 0$, which falsely states that $\rho_0(x,t)$ is completely determined given $u_p(x,t)$ (corresponding to the so-called steady-state scaling~\cite{maes_subm_2022}).

Knowing that the $\varepsilon^1$-equation is solvable, the next step is to solve it. Inverting the charge-exchange collision operator can only be done using a pseudo-inverse on the range of the operator~\cite{hutson_applications_1980, othmer_diffusion_2000, maes_subm_2022}. This requires the unknown $f_1(x,\tilde{v},t)$ to lie in the range of the charge-exchange collision operator, i.e., $f_1(x,\tilde{v},t)$ has to be a zero mean function. The implication is that the whole particle density has to be included in the equilibrium term of the Hilbert expansion: $\rho_0(x,t) = \int f_0(x,\tilde{v},t) d\tilde{v} = \int f(x,\tilde{v},t) d\tilde{v} = \rho(x,t)$. This allows for $\int f_k(x,\tilde{v},t) d\tilde{v} \equiv 0,$ $\forall k > 0$. With $f_1(x,\tilde{v},t)$ a zero mean function, the $\varepsilon^1$-equation becomes:
\begin{equation}
\partial_x \left(\tilde{c}(x,t) \rho(x,t) M(\tilde{v} \mid x,t) \right) =
-\tilde{R}_{cx}(x,t) f_1(x,\tilde{v},t),
\end{equation}
from which the first order perturbation follows as
\begin{equation}
f_1(x,\tilde{v},t) = - \frac{1}{\tilde{R}_{cx}(x,t)} \partial_x \left(\tilde{c}(x,t) \rho(x,t) M(\tilde{v} \mid x,t) \right).
\label{eq:f1_Diffusive}
\end{equation}
Integrating this expression indeed shows that $\int f_1(x,\tilde{v},t) d\tilde{v} \equiv 0$ holds, as it should.

The next step in the derivation is to solve the $\varepsilon^2$-equation. The solvability condition, using the found expressions for $f_0(x,\tilde{v},t)$ and $f_1(x,\tilde{v},t)$, results in the following constraint:
\begin{equation}
\begin{split}
&\int \partial_t (\rho(x,t) M(\tilde{v} \mid x,t)) + \partial_x \left( u_p(x,t) \rho(x,t) M(\tilde{v} \mid x,t) \right) \\
& \hspace{0.5em} -\partial_x \left( \frac{\tilde{c}(x,t)}{\tilde{R}_{cx}(x,t)} \partial_x \left(\tilde{c}(x,t) \rho(x,t) M(\tilde{v} \mid x,t) \right) \right) d\tilde{v} = 0.
\end{split}
\end{equation}
Working out the integral, the following evolution equation is obtained:
\begin{equation}
\partial_t \rho(x,t) + \partial_x(u_p(x,t)\rho(x,t)) - \partial_x \left( \frac{1}{R_{cx}(x,t)} \partial_x \left(\sigma_p^2(x,t)\rho(x,t) \right) \right) = 0,
\label{eq:diffusive_macro_eq}
\end{equation}
where the scalings of $\tilde{R}_{cx}(x,t)$ and $\tilde{\sigma}_p^2(x,t)$ cancel out such that the equation does not depend on $\varepsilon$ anymore, making this a suitable equation for computations.
Equation~\eqref{eq:diffusive_macro_eq} governs the evolution of $\rho(x,t)$ and is of advection-diffusion type. Recall that $u_p(x,t)$ and $\sigma_p^2(x,t)$ are known from the background plasma, so we have one equation for one unknown, i.e., a closed fluid model. The shape of $f_2(x,\tilde{v},t)$ follows directly from the $\varepsilon^2$-equation. In this paper, however, we settle for a first order Hilbert expansion of the form $f(x,\tilde{v},t) \approx f_0(x,\tilde{v},t) + \varepsilon f_1(x,\tilde{v},t)$ and set $f_2(x,\tilde{v},t) \equiv 0$.

\textbf{Remark 3:} Note that it is not possible to go beyond a second order perturbation. The solvability conditions for higher order perturbations $f_k(x,\tilde{v},t)$ with $k > 2$ lead to the introduction of additional evolution equations. Since the only unknown in the Hilbert expansion is $\rho(x,t)$, this would result in multiple incompatible evolution equations for the particle density. 
In the case of isotropic collision operators~\cite{othmer_diffusion_2000, othmer_diffusion_2002}, the solvability conditions are fulfilled automatically, allowing for higher order Hilbert expansions. Here, however, the drifting Maxwellian ($u_p(x,t) \neq 0$) causes the charge-exchange collision operator to be anisotropic, leading to a (maximum) second order Hilbert expansion $f(x,\tilde{v},t) = f_0(x,\tilde{v},t) + \varepsilon f_1(x,\tilde{v},t) + \varepsilon^2 f_2(x,\tilde{v},t) + \mathcal{O}(\varepsilon^3)$, that is asymptotically consistent with an error $\mathcal{O}(\varepsilon^3)$ that disappears in the limit $\varepsilon \rightarrow 0$. Note that in Ref.~\cite{othmer_diffusion_2002}, when the focus shifts away from isotropic collision operators, the authors also limit themselves to second order Hilbert expansions.

Reverting back to unscaled variables, we find the following diffusive scaling approximation to the particle velocity distribution:
\begin{equation}
\begin{split}
f(x,v,t) \approx& f_0(x,v,t) + \varepsilon f_1(x,v,t)\\[2mm]
=& \rho(x,t)M(v \mid x,t) - \frac{1}{R_{cx}(x,t)} \partial_x(c(x,t)\rho(x,t)M(v \mid x,t)).
\end{split}
\label{eq:HE_Approx_Diffusive}
\end{equation}
This approximation can be used to compute a variety of velocity dependent QoIs.

\subsubsection{Momentum and energy density}
\label{sec: mom and ene diffusive}
Expressions for the momentum and energy density can be obtained by simply using their definitions~\eqref{eq:QoIs} and inserting the first order Hilbert expansion~\eqref{eq:HE_Approx_Diffusive}. For the momentum we find:
\begin{equation}
\begin{split}
m(x,t) &\coloneqq \int vf(x,v,t) dv \\
&\approx \int v (f_0(x,v,t) + \varepsilon f_1(x,v,t))dv \\
&= \underbrace{\vphantom{\frac{1}{R_{cx}(x,t)}}\rho(x,t)u_p(x,t)}_{\substack{\text{equilibrium}}}-\underbrace{\frac{1}{R_{cx}(x,t)}\partial_x(\sigma_p^2(x,t)\rho(x,t))}_{\substack{\text{perturbation}}}.
\end{split}
\label{eq:diffusive_HE_momentum}
\end{equation}
As the scalings of $R_{cx}(x,t)$ and $\sigma_p^2(x,t)$ neutralize each other, it turns out that the equilibrium and perturbation contribution to the momentum are of equal importance.
For the energy we find:
\begin{equation}
\begin{split}
E(x,t) &\coloneqq \int \frac{v^2}{2}f(x,v,t) dv \\
&\approx \int \frac{v^2}{2} (f_0(x,v,t) + \varepsilon f_1(x,v,t))dv \\
&= \underbrace{\vphantom{\frac{1}{R_{cx}(x,t)}}\frac{1}{2}\left(\sigma_p^2(x,t)+u_p(x,t)^2 \right)\rho(x,t)}_{\substack{\text{equilibrium}}} - \underbrace{\vphantom{\frac{1}{R_{cx}(x,t)}}\frac{1}{R_{cx}(x,t)}\partial_x(u_p(x,t)\sigma_p^2(x,t)\rho(x,t))}_{\substack{\text{perturbation}}}.
\end{split}
\label{eq:diffusive_HE_energy}
\end{equation}
For the energy, the equilibrium contribution becomes increasingly more important with respect to the perturbation contribution for $\varepsilon \rightarrow 0$ (because of the increasing value of $\sigma_p^2(x,t)$). Other velocity dependent QoIs can be computed analogously.

\subsubsection{Error estimate}
\label{subsec:errorDiff}
For the first order Hilbert expansion $f(x,\tilde{v},t) \approx f_0(x,\tilde{v},t) + \varepsilon f_1(x,\tilde{v},t)$, we can add second and third order error terms: $f(x,\tilde{v},t) \approx f_0(x,\tilde{v},t) + \varepsilon f_1(x,\tilde{v},t) + \varepsilon^2 r_2(x,\tilde{v},t) + \varepsilon^3 r_3(x,\tilde{v},t)$. Inserting this into the kinetic equation~\eqref{eq:conservation_equation} and elaborating the expansion, we find that under the made assumptions $r_2(x,\tilde{v},t) = f_2(x,\tilde{v},t)$, because of the second order solvability condition (using~\eqref{eq:diffusive_macro_eq} as evolution equation). Consequently, $r_2(x,\tilde{v},t)$ has a zero particle density. The error term $r_3(x,\tilde{v},t)$ on the other hand has a non-zero particle density as it does not fulfil any solvability condition. The fluid model will therefore be correct up to $\mathcal{O}(\varepsilon^3)$ for the density $\rho(x,t)$. For the momentum, energy and other velocity dependent QoIs, the fluid model is only correct up to $\mathcal{O}(\varepsilon^2)$ because the second order error $r_2(x, \tilde{v}, t)$ has non-zero higher order velocity moments. 

Adding $f_2(x,\tilde{v},t)$ to the Hilbert expansion, the error on velocity dependent QoIs would also be $\mathcal{O}(\varepsilon^3)$. Recall that it is not possible to solve the third or higher order solvability conditions for kinetic equations dominated by an anisotropic linear collision operator (Remark 3), meaning that having errors $\mathcal{O}(\varepsilon^3)$ is the best that can be done with a diffusive scaling Hilbert expansion. For kinetic equations dominated by an isotropic linear collision operator, an arbitrary order in $\varepsilon$ can be obtained~\cite{othmer_diffusion_2000}.

In conclusion, obtaining errors $\mathcal{O}(\varepsilon^3)$ is the best that can be done for kinetic equations with diffusive scaling, dominated by an anisotropic linear collision operator, such as the charge-exchange collision operator considered in this paper.

\subsection{Hydrodynamic scaling Hilbert expansion}
\label{subsec:Hydrodynamic_scaling}
In this section, we derive a Hilbert expansion based fluid model in the so-called hydrodynamic scaling. As all the steps are analogous to those for the diffusive scaling, we will be brief here. First, we introduce the hydrodynamic scaling. There is only one physical assumption: 
\begin{itemize}
\item Collisions dominate the kinetic equation. This assumption is introduced in the kinetic equation by stating that the collision rate $R_{cx}(x,t) \sim \mathcal{O}(1/\varepsilon)$, and setting all other quantities $\sim \mathcal{O}(1)$. We introduce the scaled collision rate: $\tilde{R}_{cx}(x,t) = \varepsilon R_{cx}(x,t)$.
\end{itemize}
With this scaling, the mean free path of the particles goes to zero in the limit $\varepsilon \rightarrow 0$, as the mean free path $\lambda \sim v/R_{cx}(x,t) \sim \mathcal{O}(\varepsilon)$. Inserting the hydrodynamic scaling into the kinetic conservation equation~\eqref{eq:conservation_equation} and multiplying by $\varepsilon$ results in:
\begin{equation}
\begin{split}
&\varepsilon \partial_t f(x,v,t) + \varepsilon v \partial_x f(x,v,t) = \tilde{R}_{cx}(x,t)\\
&\times \left(M(v \mid x,t) \int f(x,v',t)dv' - f(x,v,t) \right).
\end{split}
\label{eq:Hydrodynamic_HE_scaled_kinetic_equation}
\end{equation}
We insert a Hilbert expansion~\eqref{eq:Hilbert_expansion_definition} in the scaled kinetic equation~\eqref{eq:Hydrodynamic_HE_scaled_kinetic_equation} and write equations per order in $\varepsilon$:
\begin{equation}
\begin{split}
\varepsilon^0:& \hspace{0.5em} 0 = \tilde{R}_{cx}(x,t) \left( M(v \mid x,t)\int f_0(x,v',t)dv' - f_0(x,v,t) \right)\\
\varepsilon^1:& \hspace{0.5em}  \partial_t f_0(x,v,t) + v \partial_x \left(f_0(x,v,t)\right) \\
& \quad = \tilde{R}_{cx}(x,t) \left( M(v \mid x,t)\int f_1(x,v',t)dv' - f_1(x,v,t) \right)\\
\vdots\\
\varepsilon^k:& \hspace{0.5em} \partial_t f_{k-1}(x,v,t) + v \partial_x \left(f_{k-1}(x,v,t) \right) \\
& \quad = \tilde{R}_{cx}(x,t) \left( M(v \mid x,t)\int f_k(x,v',t)dv' - f_k(x,v,t) \right).
\end{split}
\end{equation}
We first solve the $\varepsilon^0$-equation. Again defining the whole particle density to be in the equilibrium function: $\rho(x,t) \coloneqq \int f_0(x,v,t) dv$, the equilibrium follows from a straightforward manipulation:
\begin{equation}
f_0(x,v,t) = \rho(x,t)M(v \mid x,t).
\label{eq:f0_Hydrodynamical}
\end{equation}

The first order perturbation is obtained by inserting~\eqref{eq:f0_Hydrodynamical} for $f_0(x,v,t)$ in the $\varepsilon^1$-equation:
\begin{equation}
\begin{split}
    &\partial_t (\rho(x,t)M(v \mid x,t)) + v \partial_x \left(\rho(x,t)M(v \mid x,t)\right) \\
    & \hspace{0.5em} = \tilde{R}_{cx}(x,t) \left( M(v \mid x,t)\int f_1(x,v',t)dv' - f_1(x,v,t) \right).
\end{split}
\end{equation}
The solvability condition~\eqref{eq:solvability_BGK-likeCollisionOperator} for this inhomogeneous integral equation results in the following constraint:
\begin{equation}
\begin{split}
\int \partial_t (\rho(x,t)M(v \mid x,t)) + v \partial_x \left(\rho(x,t)M(v \mid x,t)\right) dv &= 0\\
\Rightarrow \hspace{2.3cm} \partial_t(\rho(x,t)) + \partial_x(u_p(x,t)\rho(x,t)) &= 0.
\end{split}
\label{eq:Hydrodynamical_HE_macro_eq}
\end{equation}
The solvability condition is the conservation of particles in the equilibrium part $f_0(x,v,t)$ of the Hilbert expansion, where advection happens with the mean velocity $u_p(x,t)$ of the Maxwellian describing the background plasma. This is an evolution equation and since we only have one degree of freedom in the equilibrium, the particle density $\rho(x,t)$, this evolution equation constitutes a closed fluid model for the hydrodynamic scaling Hilbert expansion of the conservation equation~\eqref{eq:conservation_equation}.

Since the whole particle density is in the equilibrium, we again have $\int f_k(x,v,t) dv \equiv 0,$ $\forall k > 0$, resulting in the following expression for the first order perturbation:
\begin{equation}
\begin{split}
f_1(x,v,t) =& -\frac{1}{\tilde{R}_{cx}(x,t)} \bigg( \partial_t (f_0(x,v,t)) + v\partial_x (f_0(x,v,t)) \bigg)\\
=& -\frac{1}{\varepsilon R_{cx}(x,t)} \bigg( \partial_t (\rho(x,t)M(v \mid x,t)) + v\partial_x (\rho(x,t)M(v \mid x,t)) \bigg).
\end{split}
\label{eq:f1_Hydrodynamical}
\end{equation}

We cannot go beyond a first order Hilbert expansion as higher order solvability conditions result in evolution equations that are incompatible with~\eqref{eq:Hydrodynamical_HE_macro_eq}.

\textbf{Remark 4:} In the long time limit with $t \sim \mathcal{O}(1/\varepsilon)$, the hydrodynamic evolution equation~\eqref{eq:Hydrodynamical_HE_macro_eq} will reach an equilibrium. Inserting this time scaling into the hydrodynamic scaling kinetic equation results in the so-called steady-state scaling~\cite{maes_subm_2022}.

\subsubsection{Momentum and energy density}
\label{sec: mom and ene hydro}
The first order perturbation contains no particle density, but it does contain momentum and energy. These quantities can be calculated based on the first and second moment of $\varepsilon f_1(x,v,t)$, where $f_1(x,v,t)$ is given by~\eqref{eq:f1_Hydrodynamical}:
\begin{equation}
\begin{split}
m_1(x,t) \coloneqq \int \varepsilon v f_1(x,v,t) dv =& -\frac{1}{R_{cx}(x,t)} \bigg( \partial_t (u_p(x,t)\rho(x,t))  \phantom{aaaaaaaaaa}\\
& + \partial_x \left( \left(u_p(x,t)^2 + \sigma_p^2(x,t) \right) \rho(x,t) \right) \bigg),
\end{split}
\label{eq:Hydrodynamical_HE_firstorder_mom}
\end{equation}
\begin{equation}
\begin{split}
E_1(x,t) \coloneqq \int \varepsilon \frac{v^2}{2} f_1(x,v,t) dv =& - \frac{1}{2}\frac{1}{R_{cx}(x,t)} \bigg( \partial_t \left( \left(u_p(x,t)^2 + \sigma_p^2(x,t) \right) \rho(x,t) \right) \\
& + \partial_x \left( \left(u_p(x,t)^3 + 3 u_p(x,t) \sigma_p^2(x,t) \right) \rho(x,t) \right) \bigg).
\end{split}
\label{eq:Hydrodynamical_HE_firstorder_ene}
\end{equation}
Note that these perturbations decrease in magnitude for $\varepsilon \rightarrow 0$. Also note that the expressions seemingly contain the Euler momentum and energy conservation equations, which might give the impression that the first order perturbation does not contain any momentum or energy. This is, however, not the case as momentum and energy conservation are incompatible with the evolution equation~\eqref{eq:Hydrodynamical_HE_macro_eq} for known $u_p(x,t)$ and $\sigma^2_p(x,t)$. The physical reason why there is no conservation of momentum and energy within the population of the neutral particles is that these quantities are exchanged with the plasma background (see Appendix~\ref{sec: app: momene exchange}). 

To obtain the full expressions for the momentum and energy density of the hydrodynamic scaling Hilbert expansion, the first order perturbation contributions should be added to the equilibrium contributions. The equilibrium contributions are the same as for the diffusive scaling, see~\eqref{eq:diffusive_HE_momentum} and~\eqref{eq:diffusive_HE_energy}.

\subsubsection{Error estimate}
\label{subsec:errorHydro}
Solvability conditions for higher order perturbations $f_k(x,v,t)$ with $k \geq 2$ lead to incompatible evolution equations for the particle density $\rho(x,t)$. This means that a first order hydrodynamic scaling Hilbert expansion $f(x,v,t) \approx f_0(x,v,t) + \varepsilon f_1(x,v,t)$ is the best that can be done. Under the made assumptions, we conclude that the error will be of order $\mathcal{O}(\varepsilon^2)$.

\section{Fluid model implementation}
\label{sec: solution_procedure}
In the previous sections, the kinetic equation describing neutral particle behaviour~\eqref{eq:kinetic_equation} has been split in a source, conservation, and ionization equation~\eqref{eq:source_equation}-\eqref{eq:sink_equation}. The fluid model equivalents for the source and ionization equation are given by~\eqref{eq:moments_of_source} and~\eqref{eq:moments_of_i}, respectively. For the conservation equation, Hilbert expansion based fluid models are derived in the diffusive scaling~\eqref{eq:diffusive_macro_eq},~\eqref{eq:diffusive_HE_momentum},~\eqref{eq:diffusive_HE_energy} and in the hydrodynamic scaling~\eqref{eq:Hydrodynamical_HE_macro_eq},~\eqref{eq:Hydrodynamical_HE_firstorder_mom},~\eqref{eq:Hydrodynamical_HE_firstorder_ene}. Recall that~\eqref{eq:Hydrodynamical_HE_firstorder_mom} and~\eqref{eq:Hydrodynamical_HE_firstorder_ene} only contain the perturbation contributions. Obtaining the full momentum and energy expressions requires adding the equilibrium parts of~\eqref{eq:diffusive_HE_momentum} and~\eqref{eq:diffusive_HE_energy}, respectively. 

This section deals with solving the resulting transient fluid models using a time stepping procedure. First, the different equations are discretized for the numerical experiments in Section~\ref{sec: experiments}. Next, the estimation of macroscopic QoIs is discussed for the fluid models, a discrete velocity model, and a particle tracing Monte Carlo method, the last two being reference solvers used for the numerical experiments in Section~\ref{sec: experiments}. Finally, the algorithmic solution procedures for the fluid models and reference solvers are described.

\subsection{Discretization of the fluid models}
Implementing the fluid model equivalents of the split equations~\eqref{eq:source_equation}-\eqref{eq:sink_equation} requires choosing a discretization for the different equations. There are no spatial derivatives in the source and ionization equation. Their solution, as provided in Section~\ref{sec:bulk}, is readily implemented. For the conservation equation, there are spatial dependencies that have to be discretized.

The two Hilbert expansion based fluid models (diffusive and hydrodynamic scaling) can be written in the following form:
\begin{equation}
    \underbrace{\partial_t \rho(x,t)}_{\substack{\text{transient}}} + \underbrace{\partial_x (u_p(x,t) \rho(x,t))}_{\substack{\text{advection}}} - \underbrace{\partial_x (D(x,t) \partial_x (\sigma_p^2(x,t) \rho(x,t)) )}_{\substack{\text{diffusion}}} = 0,
    \label{eq:ADP_Fluid_Eq}
\end{equation}
with $D(x,t) = \frac{1}{R_{cx}(x,t)}$ in the diffusive scaling~\eqref{eq:diffusive_macro_eq} and $D(x,t) = 0$ in the hydrodynamic scaling~\eqref{eq:Hydrodynamical_HE_macro_eq}. The two fluid models can therefore be modelled using the same simulation code by only changing the diffusion coefficient $D(x,t)$.

The equation is semi-discretized using a finite volume scheme on a grid with uniform spacing $\Delta x$. Cell centers are located at the positions $x_j$, $j=1,\hdots,J$, where $J$ is the total number of grid cells. The advection term is discretized using a first order upwind scheme, the diffusion term using a second order centered discretization. Boundaries are treated periodically. The semi-discretized equation can be written in condensed form:
\begin{equation}
    \partial_t \rho(x_j,t) + \Phi(x_j,t) = 0,
\end{equation}
where $\Phi(x_j,t)$ represents the spatially discretized advection and diffusion terms at position $x_j$. The time discretization is done using explicit Euler with time step $\Delta t$:
\begin{equation}
    \rho(x_j,t_i+\Delta t) \approx \rho(x_j,t_i) - \Delta t \Phi(x_j,t_i).
    \label{eq:ExplicitEulerDiscretization}
\end{equation}

At each time in the simulation, the momentum and energy density of the Hilbert expansion based fluid models can be calculated by discretizing the formulas discussed in Section~\ref{sec: mom and ene diffusive} and~\ref{sec: mom and ene hydro}, using the same discretization schemes as for the evolution equation~\eqref{eq:ADP_Fluid_Eq}. Other QoIs can be treated analogously.

\subsection{Estimation of macroscopic quantities of interest}
Transient simulations run from an initial time $t=0$ up to a given time $t_1$ at which the QoIs are to be computed. The QoIs at time $t_1$ can be obtained from a Hilbert expansion based fluid model simulation in a straightforward way by inserting the Hilbert expansion~\eqref{eq:Hilbert_expansion_definition} in the velocity integrals~\eqref{eq:QoI_Generic} defining the QoIs. 

In the numerical experiments in Section~\ref{sec: experiments}, the Hilbert expansion based fluid models are compared to a discrete velocity model~\cite{mieussens_discrete_2000} and a particle tracing Monte Carlo method~\cite{spanier_monte_1969, mortier_advanced_2020, lapeyre_introduction_2003}. Therefore, the QoIs also have to be computed by these two reference solvers. For discrete velocity models, calculating QoIs is a straightforward process, as an approximation to the particle velocity distribution $f(x,v,t)$ is readily available. For particle tracing Monte Carlo methods, obtaining an accurate solution at a fixed point in time requires an extremely large number of particles because of two reasons. First, a fraction of the initialized particles does not reach the time $t_1$ due to ionization processes (and in a more general setting other potential particle sinks such as absorption at a boundary). Second, the particles that survive up to time $t_1$ only contribute once to the estimates of the QoIs. These two effects combined result in a high variance on the Monte Carlo estimates of the QoIs.

To obtain good reference solutions at a reasonable computational cost, we reduce the variance on the Monte Carlo estimation by not asking for the QoIs at a given point $t_1$ in time, but by asking for the average QoIs over a time window $W$: $t \in W = [t_1, t_2]$: 
\begin{equation}
        \bar{Q}(x_j) = \frac{1}{\Delta_W}\int_{t_1}^{t_2}\int_V q(v) f(x_j,v,t) dvdt,
    \label{eq:time_window_estimators_MC}
\end{equation}
where $\Delta_W = t_2-t_1$ represents the width of the time window. In a particle tracing Monte Carlo method, this type of integral is a standard output that can be estimated using a time-integrated estimator over $W$, e.g., using an analog or track-length estimator~\cite{mortier_advanced_2020, spanier_monte_1969}. In time-integrated estimators, the different particles contribute multiple times to the estimates of the QoIs during the time window $W$, reducing the variance on the results. 

For the fluid solvers and the discrete velocity model, the solution can also be integrated over the time window $W$ and divided by $\Delta_W$ to obtain the average solution over the time window. Assuming that the calculated QoIs are constant in each time step $\Delta t$ of the simulation and choosing $\Delta t$ such that $\Delta_W = N\Delta t$ with $N \in \mathbb{N}$, the averaged QoIs can be approximated as:
\begin{equation}
        \bar{Q}(x_j) \approx \frac{1}{\Delta_W}\sum_{n=1}^N \Delta t Q_n(x_j),
    \label{eq:time_window_estimators_fluid}
\end{equation}
where $Q_n(x_j)$ represents the QoIs in time step $n$ and is calculated by evaluating the corresponding velocity integrals~\eqref{eq:QoI_Generic} in that time step.

\subsection{Solution procedure}
Simulating the Hilbert expansion based fluid models requires solving the fluid model equivalents of the split equations~\eqref{eq:source_equation}-\eqref{eq:sink_equation} sequentially in each time step. For the employed first order splitting, the order in which the equations are solved is not important. We first solve the source equation, then the Hilbert expansion based conservation equation and finally the ionization equation, following the natural path of neutral particles from source to sink, as is done in particle tracing Monte Carlo methods~\cite{mortier_advanced_2020, spanier_monte_1969, lapeyre_introduction_2003}. The source equation~\eqref{eq:source_equation} only has to be solved for the particle density $\rho(x,t)$. After solving the source equation, the particle density is evolved using the Hilbert expansion based conservation equation. Next, the QoIs are calculated by inserting the Hilbert expansion ansatz in the velocity integrals~\eqref{eq:QoI_Generic}. Finally, the ionization equation is solved in which a fraction of the QoIs is lost in the ionization sink.

Hilbert expansion based fluid model simulations start with the construction of the initial condition. Then a time stepping procedure with a time step $\Delta t$ is executed up to time $t_1$: the beginning of the estimation time window. Next, the simulation is continued until time $t_2$ while counting contributions to the QoIs, averaged over the estimation time window $W$. The pseudo-code is given in Algorithm~\ref{alg:HEFluidModel} in Appendix~\ref{sec: app: pseudo-code}.

We also briefly outline the solution procedure for the two reference solvers used in Section~\ref{sec: experiments}. Discrete velocity models directly discretize the kinetic equation on a $x,v$-grid in phase space. The same equidistant space and time discretizations as used for the fluid models are employed. The velocity is discretized using Gauss-Hermite points~\cite{liu_note_1994}, based on a Maxwellian with mean velocity $\frac{1}{J}\sum^J_{j=1}u_p(x_j,t=0)$ and variance $\frac{1}{J}\sum^J_{j=1}\sigma^2_p(x_j,t=0)$. For more details on discrete velocity models, we refer the reader to Ref.~\cite{mieussens_discrete_2000}. Transient particle tracing Monte Carlo methods~\cite{lapeyre_introduction_2003, reiter_time_1995} simulate particle trajectories in phase space. To count contributions to the QoIs, the same finite volume $x$-grid as used for the fluid models can be used to build a histogram. In the experiments, a stationary source $S(x,v,t) = S(x,v)$ will be assumed, allowing for the introduction of particles in the simulation by sampling the generation time $t^*$ of each particle from a uniform distribution over the time interval $[0, t_2]$. There is no need to sample times beyond $t_2$ as those particles will never contribute to the estimators. Each particle is tracked from time $t^*$ to the end of the estimation time window $t_2$ or until it is lost in a sink. Pseudo-code for such a particle tracing Monte Carlo method is given in Algorithm~\ref{alg:transientMC_StationarySource} in Appendix~\ref{sec: app: pseudo-code}.

\section{Numerical experiments}
\label{sec: experiments}
In this section, we test the performance of the two Hilbert expansion based fluid models in three numerical experiments. In the first experiment, we set up a test case in the hydrodynamic scaling and in the diffusive scaling, applying the two Hilbert expansion based fluid models to each test case. The second experiment compares the diffusive scaling Hilbert expansion based fluid model to a phenomenological pressure-diffusion model~\cite{horsten_fluid_2019, horsten_comparison_2016}, which is derived based on a variant of the Method of Moments combined with physical intuition based assumptions. In the third experiment, the two Hilbert expansion based fluid models are applied to a test case with parameter values in the realistic ranges for fusion reactor plasma edge modelling. The test case comprises a fictitious gas puffing event in the plasma edge, where the injected neutral particles instantaneously equilibrate with the background plasma. 

For each test case, a reference solution is computed using a particle tracing Monte Carlo method with analog simulation and track-length estimation~\cite{mortier_advanced_2020, spanier_monte_1969}, and using a discrete velocity model with 200 Gauss-Hermite points to discretize the velocity. The Monte Carlo method is seeded to ensure reproducibility of the results. More details on the reference solutions can be found in Appendix~\ref{sec: app: errors}. Each experiment is performed in the domain $x \in [0,1]$, which is subdivided uniformly in 200 finite volume grid cells. 

The two Hilbert expansion based fluid models use the same time step within a given experiment. The time step restriction in the hydrodynamic scaling experiments is due to the advection CFL number
\begin{equation}
\text{CFL}_\text{adv} = \frac{\text{max}(u_p)\Delta t}{\Delta x},
\label{eq:CFL_ADV}
\end{equation}
and in the diffusive scaling experiments due to the (more restrictive) diffusive CFL number
\begin{equation}
\text{CFL}_\text{diff} = \frac{1}{2}\frac{\text{max}\left(\frac{\sigma_p^2}{R_{cx}}\right) \Delta t}{\Delta x^2}.
\label{eq:CFL_DIFF}
\end{equation}
For each experiment, the time step is chosen such that a CFL number equal to $0.5$ is obtained. The CFL numbers are independent of $\varepsilon$ for their respective scalings, which means that the same number of time steps is required to reach a chosen time $t_1$, no matter the value of $\varepsilon$. As a result, the cost for solving the fluid models is independent of $\varepsilon$. This is in contrast to the particle tracing Monte Carlo methods, where the cost increases dramatically for $\varepsilon \rightarrow 0$ (high-collisional regime), as each collision is executed explicitly. The Hilbert expansion based fluid models therefore provide an enormous speed-up over particle tracing Monte Carlo methods in high-collisional regimes. We refer to Appendix~\ref{sec: app: pseudo-code} for an additional remark on the cost as a function of the grid refinement. The code used to perform the experiments is openly available in \reponame.

\subsection{Experiment 1: diffusive scaling versus hydrodynamic scaling}
\subsubsection{Setup}
\label{sec:experiment1:setup}
In the first experiment, we test the two Hilbert expansion based fluid models in a hydrodynamic scaling and in a diffusive scaling test case. All the parameter profiles are chosen sinusoidal such that they are smooth and periodic. We simulate starting from the following initial condition:
\begin{equation}
\rho(x,t=0) = 1 + \frac{1}{2\pi} \sin(2 \pi x).
\end{equation}
The estimation time window starts at the initial time: $t_1 = 0$. The estimation time window has a width of 10 time steps for the hydrodynamic scaling test case and 5000 time steps for the diffusive scaling test case, respectively, such that the simulated time for both test cases is about $\Delta_W \approx 0.025$. The source is set to zero: $S(x,v,t) \equiv 0$. The collision rates $R_i(x,t)$, $R_{cx}(x,t)$ and background plasma quantities $u_p(x,t)$, $\sigma_p^2(x,t)$ are chosen to be stationary functions. The model parameter profiles for the hydrodynamic scaling test case are the following:
\begin{equation}
\begin{split}
R_i(x) &= 0,\\
R_{cx}(x) &= 10^{a} \left( 1 +  \frac{1}{4\pi} \sin(4 \pi x) \right),\\
u_p(x) &= 1 + \frac{1}{6\pi} \sin(6 \pi x),\\
\sigma_p^2(x) &= 1 + \frac{1}{6\pi} \cos(6 \pi x),
\end{split}
\end{equation}
where $a$ allows to tune the value of $\varepsilon$, as we can interpret $\varepsilon = 10^{-a}$. For the diffusive scaling test case, all the functions are kept the same, with the exception of the variance $\sigma_p^2(x)$, which becomes
\begin{equation}
\sigma_p^2(x) = 10^a \left( 1 + \frac{1}{6\pi} \cos(6 \pi x) \right).
\end{equation}
In the diffusive scaling, we can interpret $\varepsilon^2 = 10^{-a}$. For both test cases, we can set $\varepsilon \rightarrow 0$ by increasing the value of $a$.

The model parameter profiles are chosen such that they correspond either to the hydrodynamic scaling or to the diffusive scaling. However, both the Hilbert expansion based fluid models can be applied to each of the test cases. The hydrodynamic scaling fluid model assumes that $\varepsilon \sim \frac{1}{R_{cx}} \sim 10^{-a}$, which is correct in the hydrodynamic scaling test case, but not in the diffusive scaling test case. The diffusive scaling fluid model assumes that $\varepsilon^2 \sim  \frac{1}{R_{cx}},\frac{1}{\sigma_p^2} \sim 10^{-a}$, which is correct in the diffusive scaling test case, but not in the hydrodynamic scaling test case. Applying the correct Hilbert expansion based fluid model to a test case allows to assess its performance. Applying the wrong Hilbert expansion based fluid model to a test case allows to assess the influence of using a wrong scaling.

\subsubsection{Results}
We are interested in the accuracy of the Hilbert expansion based fluid models for $\varepsilon \rightarrow 0$, i.e., for increasing $a$. We perform the simulations for $a = 0, 1, \hdots, 5$ and calculate the relative errors with respect to the reference solution
\begin{equation}
\text{relative error} = \frac{1}{J}\sum^J_{j=1} \left| \frac{\bar{Q}(x_j)-\bar{Q}_{ref}(x_j)}{\bar{Q}_{ref}(x_j)} \right|
\label{eq:rel_error_calculation}
\end{equation}
for the particle, momentum, and energy density. The hydrodynamic scaling test case uses the discrete velocity model solution as reference solution, while the diffusive scaling test case uses the particle tracing Monte Carlo solution as reference solution, as explained in Appendix~\ref{sec: app: errors}.

Figure~\ref{fig:Exp1_Eps} shows that for both the hydrodynamic scaling and diffusive scaling test case, the correctly scaled fluid models converge towards the reference solutions for increasing $a$, i.e., for $\varepsilon \rightarrow 0$. However, the predicted convergence rates (Section~\ref{subsec:errorDiff} and~\ref{subsec:errorHydro}) are not achieved. This seems to indicate that for low values of $a$, the fluid models are not yet in the asymptotic regime. For high values of $a$, where calculating the reference solutions becomes increasingly expensive, the statistical and discretization errors quickly become dominant over the modelling error. This prevents a proper illustration of the asymptotic behaviour of the modelling error. Nevertheless, Figure~\ref{fig:Exp1_Eps} clearly shows that the modelling error decreases rapidly in the correct asymptotic limit.

\begin{figure}
\centering
\begin{subfigure}[t]{.48\textwidth}
  \centering
  \includegraphics[width=\linewidth]{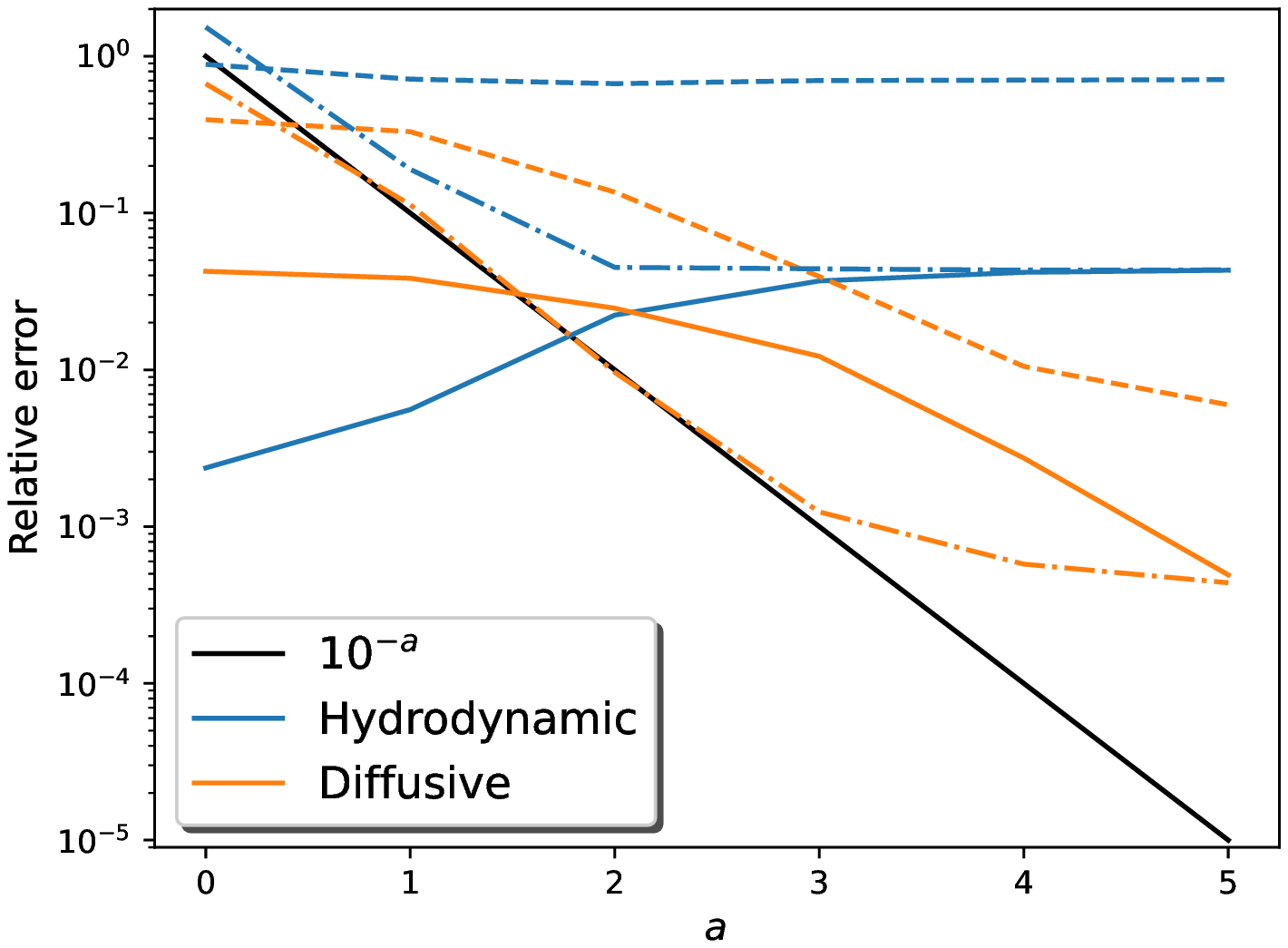}
  \caption{Diffusive scaling experiment.}
  \label{fig:sub:exp1_diff}
\end{subfigure}
\begin{subfigure}[t]{.48\textwidth}
  \centering
  \includegraphics[width=\linewidth]{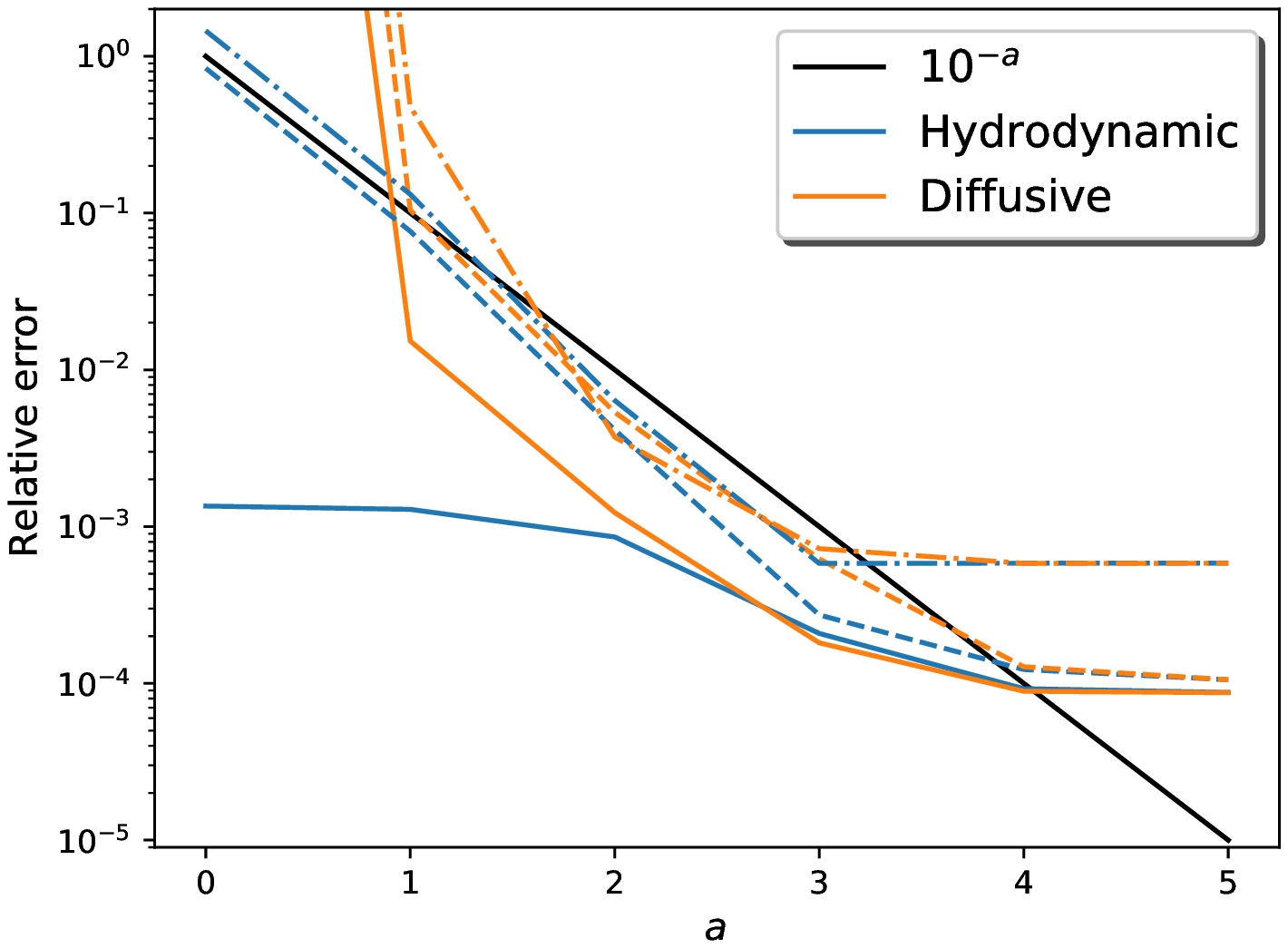}
  \caption{Hydrodynamic scaling experiment.}
  \label{fig:sub:exp1_hydro}
\end{subfigure}
\caption{Relative errors of the diffusive ($\varepsilon^2 = 10^{-a}$) and hydrodynamic ($\varepsilon = 10^{-a}$) scaling Hilbert expansion based fluid models as a function of $a$. Full lines: error on the averaged particle density $\bar{\rho}$. Dashed lines: error on the averaged momentum density $\bar{m}$. Dash-dotted lines: error on the averaged energy density $\bar{E}$.}
\label{fig:Exp1_Eps}
\end{figure}

\begin{figure}
\centering
\begin{subfigure}[t]{\textwidth}
\includegraphics[width=\linewidth]{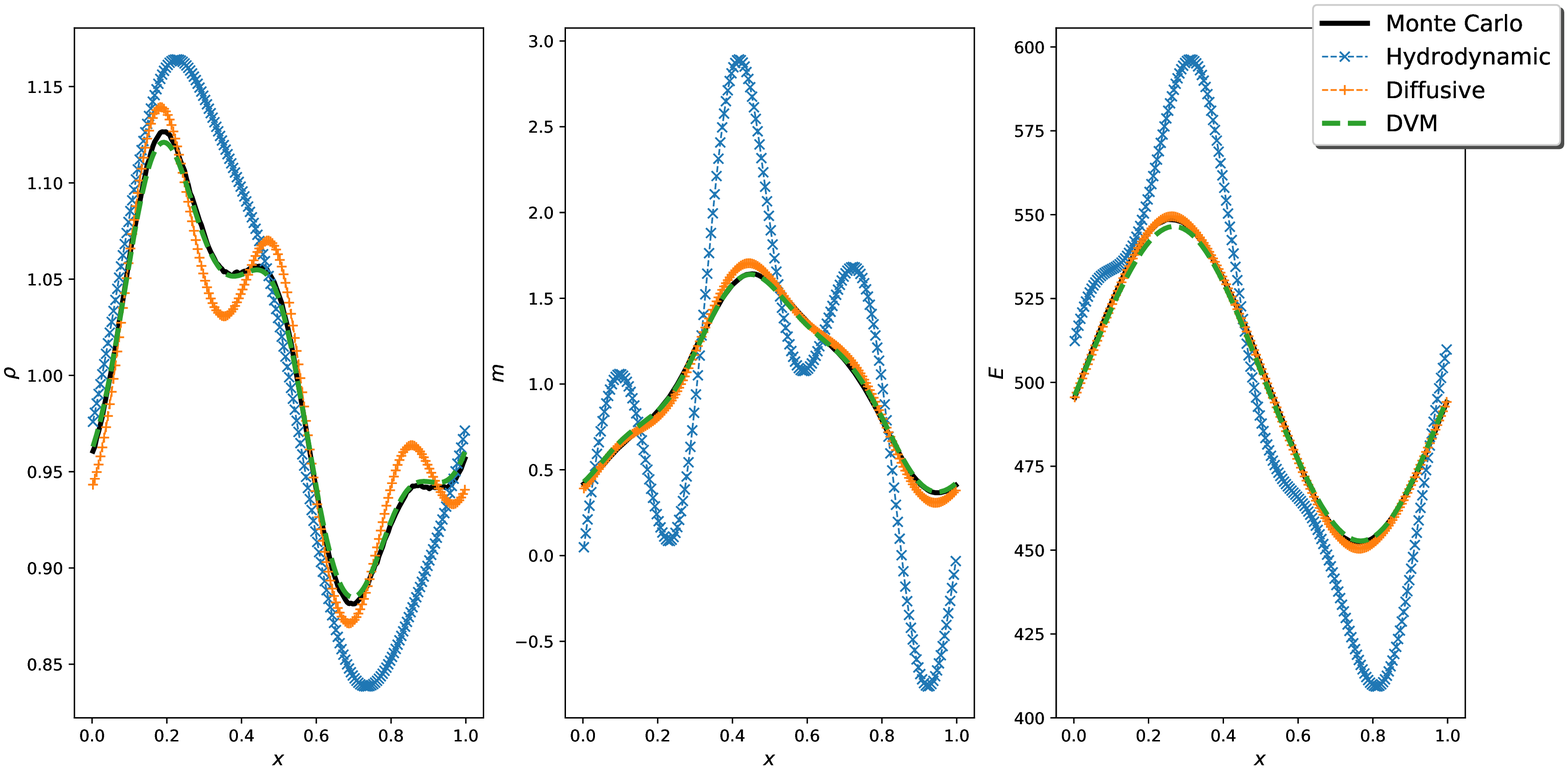}
\caption{Diffusive scaling experiment with $a=3$.}
\label{fig:sub:exp1_diff_fig2}
\end{subfigure}
\begin{subfigure}[t]{\textwidth}
\includegraphics[width=\linewidth]{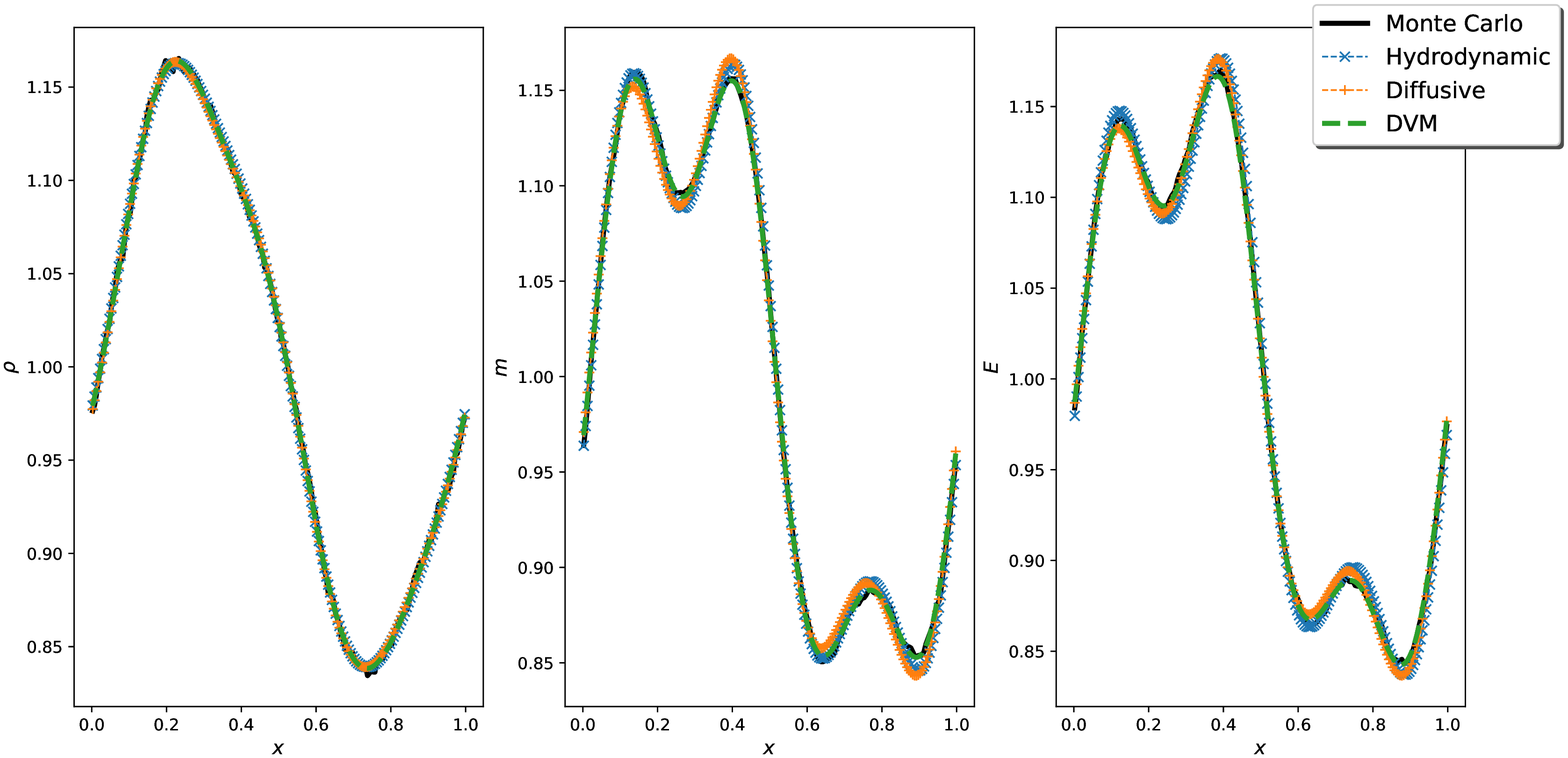}
\caption{Hydrodynamic scaling experiment with $a=2$.}
\label{fig:sub:exp1_hydro_fig2}
\end{subfigure}
\caption{Particle tracing Monte Carlo, Hilbert expansion based fluid models, and discrete velocity model (DVM) solutions. Left to right: averaged particle, momentum, and energy density.}
\label{fig:Exp1_QoIs}
\end{figure}

Looking at the diffusive scaling test case (Fig.~\ref{fig:sub:exp1_diff}), the difference between the two Hilbert expansion based fluid models is obvious. The hydrodynamic scaling fluid model has no information on the variance $\sigma^2_p(x)$ being high and therefore does not exhibit the diffusive behaviour induced by this high variance. For increasing $a$, the diffusive scaling fluid model converges towards the reference solution (apart from the introduced discretization and statistical errors), while this is not the case for the hydrodynamic scaling fluid model.

Looking at the hydrodynamic scaling test case (Fig.~\ref{fig:sub:exp1_hydro}), we see that for $a=\{0,1\}$ the forward Euler discretization of the diffusive scaling fluid model is unstable, resulting in the observed large errors. For larger values of $a$, the diffusive scaling fluid model is stable and converges to the performance of the hydrodynamic scaling fluid model. The stagnation of the errors for high values of $a$ is due to the discretization errors becoming dominant.

The discrepancies between the two fluid models and the two reference solutions are still visible for $a=3$ ($\varepsilon^2 = 10^{-3}$) in the diffusive scaling test case and $a=2$ ($\varepsilon=10^{-2}$) in the hydrodynamic scaling test case, as shown in Figure~\ref{fig:Exp1_QoIs}. In the diffusive scaling test case (Fig.~\ref{fig:sub:exp1_diff_fig2}), the lack of diffusion in the hydrodynamic scaling fluid model is clearly visible. In the hydrodynamic scaling test case (Fig.~\ref{fig:sub:exp1_hydro_fig2}), the diffusive scaling fluid model is only slightly outperformed by the hydrodynamic scaling fluid model. This can be explained by noting that in~\eqref{eq:diffusive_macro_eq} the diffusion term becomes negligible for a low variance $\sigma^2_p(x)$ combined with an increasing collision rate $R_{cx}(x)$, leading to~\eqref{eq:Hydrodynamical_HE_macro_eq} in the high-collisional limit.

The conclusions of this first experiment are (i) that the modelling errors decrease rapidly when a Hilbert expansion based fluid model is applied to a problem that has the correct scaling; (ii) that the diffusive scaling model reduces to the  hydrodynamic scaled problem if the variance $\sigma^2_p(x)$ is low. The stability condition of the diffusive scaling model, however, is more restrictive than the one of the hydrodynamic scaling model.

\subsection{Experiment 2: comparison with phenomenological pressure-diffusion model}
In the second experiment, we compare the diffusive scaling Hilbert expansion based fluid model with the phenomenological pressure-diffusion model described in Refs.~\cite{horsten_fluid_2019, horsten_comparison_2016}. The pressure-diffusion model is derived by using a variant of the Method of Moments, where closure is achieved by assuming that the temperature of the neutral particles equals the temperature of the background plasma, resulting in the constraint $\sigma^2(x) = \sigma^2_p(x)$. The resulting fluid equations are Navier-Stokes like and are reduced to a pressure-diffusion model by neglecting several terms in the Navier-Stokes like fluid model based on physical intuition.

We reuse the diffusive scaling test case from Experiment 1. In the absence of sources and sinks, only the conservation equation~\eqref{eq:conservation_equation} is solved in the Hilbert expansion based fluid model. Setting the sources and sinks to zero in the pressure-diffusion model reduces that model's pressure-diffusion equation to the evolution equation~\eqref{eq:diffusive_macro_eq} of the diffusive scaling Hilbert expansion. The momentum equation of the pressure-diffusion model corresponds to equation~\eqref{eq:diffusive_HE_momentum}. The remaining difference between the two models is the difference between their energy equations. The energy density in the pressure-diffusion model follows from the calculated momentum~\eqref{eq:diffusive_HE_momentum} and the assumption $\sigma^2(x) = \sigma_p^2(x)$:
\begin{equation}
E(x,t) = \frac{1}{2} \left(\frac{m(x,t)^2}{\rho(x,t)} + \rho(x,t)\sigma_p^2(x) \right).
\end{equation} 
The Hilbert expansion based fluid model has energy density~\eqref{eq:diffusive_HE_energy} from which a $\sigma^2(x) \neq \sigma_p^2(x)$ can be derived using the definitions below equation~\eqref{eq:QoIs}. 

\begin{figure}[t]
\centering
\begin{subfigure}[t]{.48\textwidth}
  \centering
  \includegraphics[width=\linewidth]{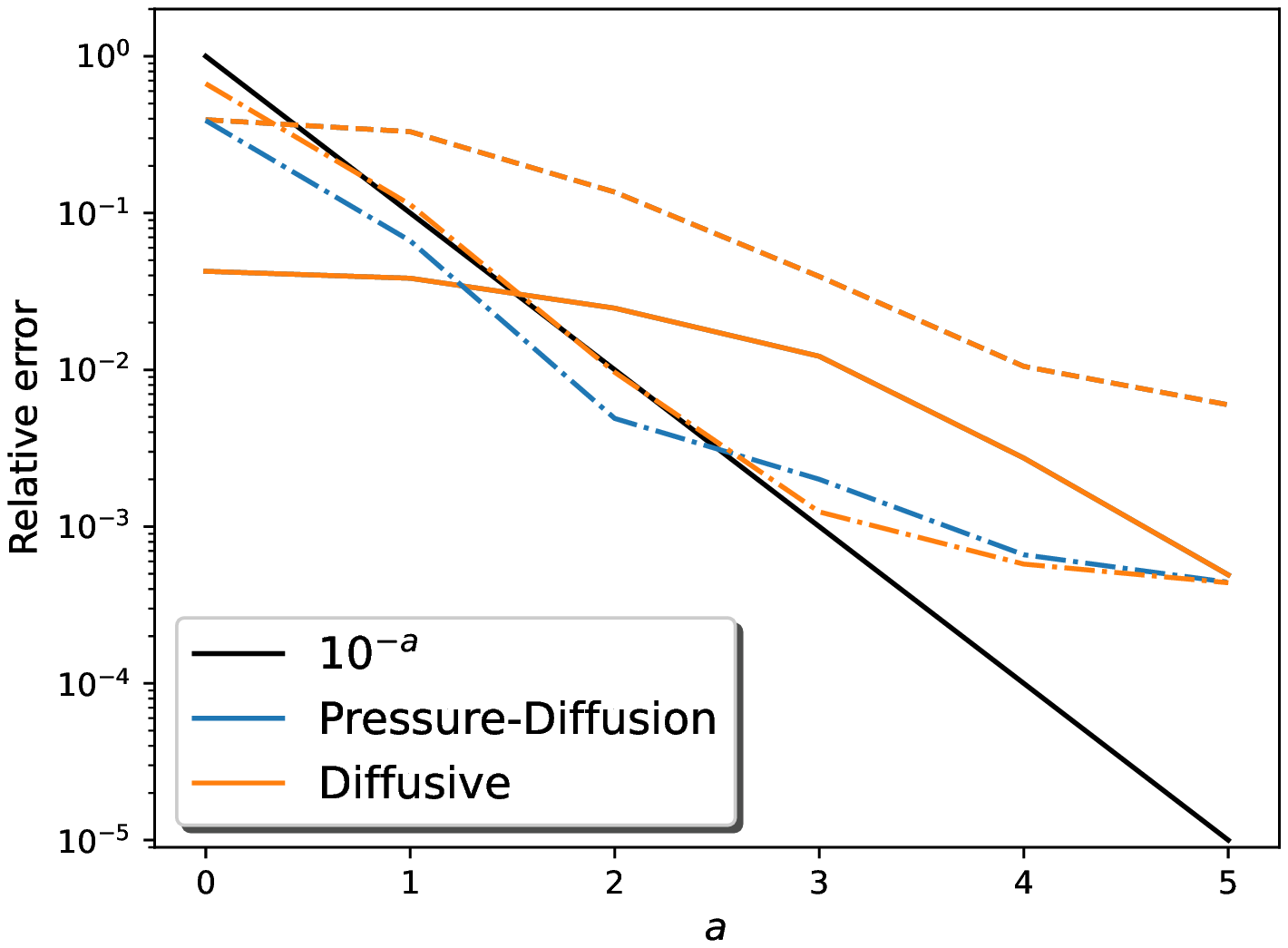}
  \caption{Relative errors for the first order diffusive scaling Hilbert expansion based fluid model and the pressure-diffusion fluid model.}
  \label{fig:sub:exp2_eps}
\end{subfigure}
\begin{subfigure}[t]{.48\textwidth}
  \centering
  \includegraphics[width=\linewidth]{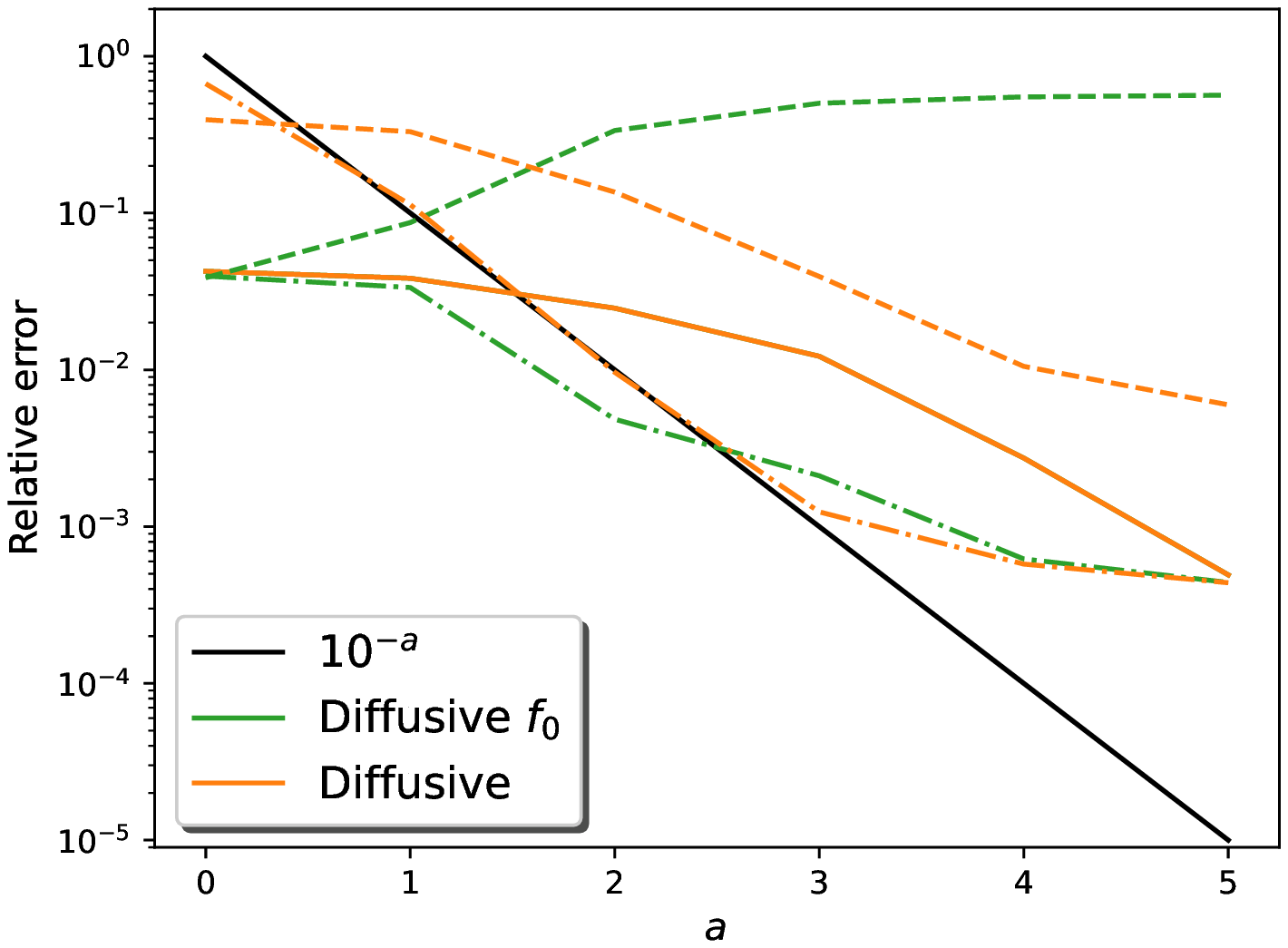}
  \caption{Relative errors for the zeroth and first order diffusive scaling Hilbert expansion based fluid model.}
  \label{fig:sub:exp2_epsDiffHE0}
\end{subfigure}
\begin{subfigure}[t]{.48\textwidth}
  \centering
  \includegraphics[width=\linewidth]{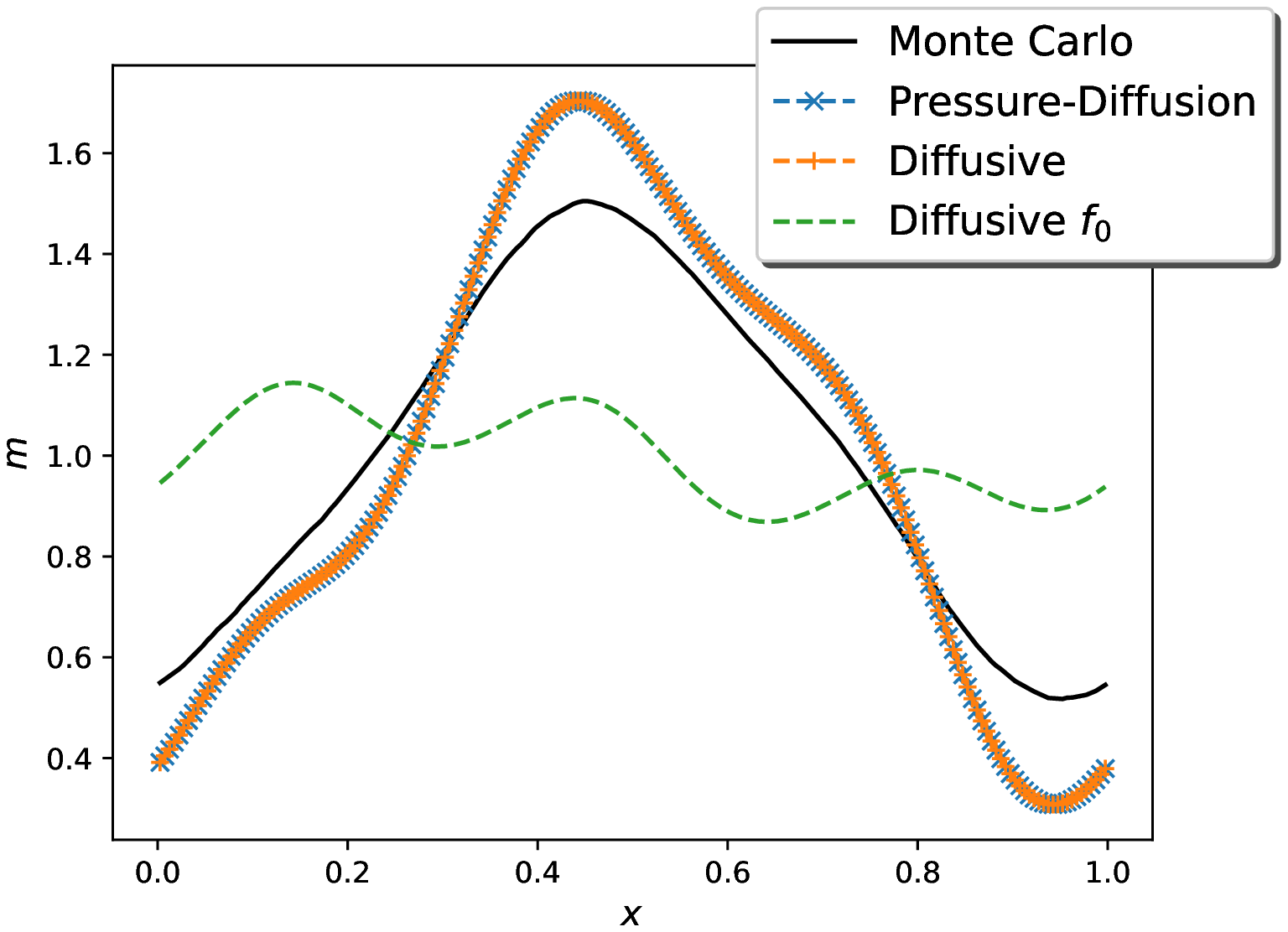}
  \caption{Momentum density $m$ for $a=2$.}
  \label{fig:sub:exp2_mom}
\end{subfigure}
\begin{subfigure}[t]{.48\textwidth}
  \centering
  \includegraphics[width=\linewidth]{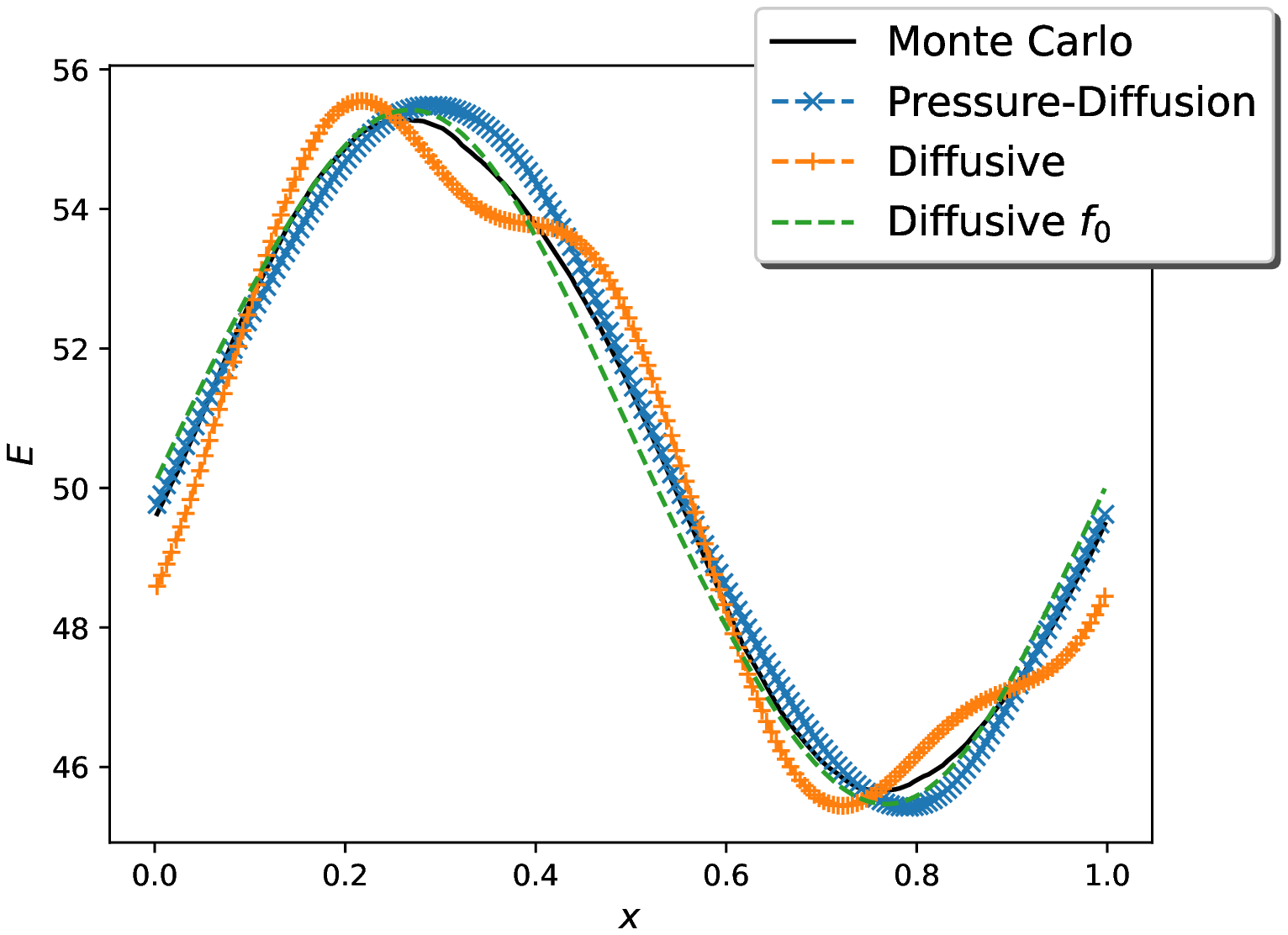}
  \caption{Energy density $E$ for $a=2$.}
  \label{fig:sub:exp2_ene}
\end{subfigure}
\caption{First and zeroth order diffusive scaling Hilbert expansion based fluid model ($\varepsilon^2 = 10^{-a}$) versus phenomenological pressure-diffusion fluid model of Refs.~\cite{horsten_fluid_2019, horsten_comparison_2016}. Top row: Relative errors as a function of $a$; full lines: error on the averaged particle density $\bar{\rho}$; dashed lines: error on the averaged momentum density $\bar{m}$; dash-dotted lines: error on the averaged energy density $\bar{E}$. Bottom row: averaged momentum and energy density for $a=2$.}
\label{fig:Exp2}
\end{figure}

The description of the similarities and differences between the two models already shows that phenomenological fluid models incorporating expert knowledge, such as the pressure-diffusion model, can come very close to models that follow from more systematic derivations, as is the case for the Hilbert expansion based fluid model. The comparison between the two models is shown in Figure~\ref{fig:Exp2}. The relative errors of the phenomenological pressure-diffusion model and diffusive scaling Hilbert expansion based fluid model as a function of $a$ are shown in Figure~\ref{fig:sub:exp2_eps}. The relative errors on the particle density $\rho(x,t)$ and momentum density $m(x,t)$ coincide, because the two models solve the same equations. For the energy density $E(x,t)$, we see that the relative errors are different, but very close to each other. Figures~\ref{fig:sub:exp2_mom} and~\ref{fig:sub:exp2_ene} respectively show the momentum density and energy density for $a=2$ ($\varepsilon^2 = 10^{-2}$). For that value of $a$, the difference in the energy density for the two models is still clearly visible.

For $a<3$, the phenomenological pressure-diffusion model slightly outperforms the Hilbert expansion based fluid model, which is also visible in Figure~\ref{fig:Exp2}. The phenomenological pressure-diffusion model turns out to be the Hilbert expansion based fluid model, but with an adapted approximation $f(x,v,t) \approx f_0^*(x,v,t)$ for calculating the energy, where $f_0^*(x,v,t)$ is a drifting Maxwellian with mean $u(x)\neq u_p(x)$ and variance $\sigma^2(x) = \sigma_p^2(x)$. The superior accuracy can then be explained by the fact that a Hilbert expansion's perturbation contribution $\varepsilon f_1(x,v,t)$ to a QoI (to the energy density) is less accurate for small values of $a$ than for large values of $a$. Far away from the asymptotic limit $\varepsilon \rightarrow 0$ (for low $a$ values), a lower order approximation (a drifting Maxwellian) is more robust than a higher order approximation (a drifting Maxwellian with first order perturbation). Note, however, that for the momentum density the perturbation contribution $\varepsilon f_1(x,v,t)$ and the equilibrium contribution $f_0(x,v,t)$ are equally important (see Section~\ref{sec: mom and ene diffusive}) and therefore the perturbation contribution cannot be neglected in order to obtain a better approximation. The insight that lower order approximations are more robust (provided that the perturbation and equilibrium are not equally important for the QoI under consideration) can readily be incorporated in the deployment of Hilbert expansion based fluid models.

To verify this insight, Figure~\ref{fig:sub:exp2_epsDiffHE0} shows the relative errors for the zeroth order diffusive scaling Hilbert expansion based fluid model, which uses $f(x,v,t) \approx f_0(x,v,t)$ as lower order approximation for the particle velocity distribution. The momentum density and energy density of this fluid model are plotted in figures~\ref{fig:sub:exp2_mom} and~\ref{fig:sub:exp2_ene}, respectively. As expected, the error on the momentum density does not decrease anymore for increasing $a$, because the perturbation contribution and the equilibrium contribution of the first order Hilbert expansion are of equal importance for the momentum density. For the energy density, it is observed that the error still decreases for increasing $a$, because the perturbation contribution becomes negligible in the asymptotic limit, i.e., using $f(x,v,t) \approx f_0(x,v,t)$ is asymptotically consistent for the energy density. Additionally, the accuracy has improved for low $a$ values ($a<3$), making the lower order approximation a better, more robust alternative to the first order Hilbert expansion. 

Even though the accuracy of the phenomenological pressure-diffusion fluid model and the Hilbert expansion based fluid models is very similar, there are two additional advantages of using the systematic Hilbert expansion based approach. The first advantage is that the Hilbert expansion based approach introduces physical assumptions simply by scaling model parameters at the kinetic description level, while in the phenomenological pressure-diffusion model the assumptions are introduced in a more ad hoc way at the fluid description level by neglecting terms and adding constraints. As a result, the scalings introduced by the Hilbert expansion based approach give a clear view on the range of validity of the resulting fluid models, as the required relative magnitude of the different model parameters is made explicit.

The second advantage is that the Hilbert expansion based approach provides direct access to the approximate particle velocity distribution underlying the resulting fluid model, while that is not the case for the pressure-diffusion model. The reason for this is that in the Hilbert expansion based approach, the assumptions are already introduced at the kinetic description level, which elucidates their influence on the particle velocity distribution during the derivation of the fluid equations. In a phenomenological fluid model, assumptions are made at the fluid description level, leaving their effect on the particle velocity distribution unclear. 

\subsection{Experiment 3: realistic application case}
In the final experiment, we set the model parameters such that they correspond to realistic plasma edge values for a pure hydrogen plasma in a tokamak. The experiment consists of modelling a fictitious gas puffing event in a plasma representative for cold, dense divertor conditions~\cite{stangeby_plasma_2000}. For this experiment, the parameters and variables are assigned units given in Table~\ref{table:units}.

\begin{table}[!ht]
    \centering
    \begin{tabular}{|l|l|l|}
    \hline
      Parameter/variable & Symbol & Units \\ \hline\hline
        Time & $t$ & $s$ \\ \hline
        Position & $x$ & $m$ \\ \hline
        Velocity & $v$, $u_p$ & $m$ $s^{-1}$ \\ \hline
        Plasma velocity variance & $\sigma_p^2$ & $(m$ $s^{-1})^2$ \\ \hline
        Temperature & $T_i$, $T_e$  & eV \\ \hline
        Electronvolt & $e$ & J $\text{eV}^{-1}$ $=$ kg $(m$ $s^{-1})^2$ $\text{eV}^{-1}$\\ \hline
        Plasma particle mass & $m_p$ & kg \\ \hline
        Collision rate & $R_i$, $R_{cx}$ &  $s^{-1}$ \\ \hline
        Particle velocity distribution & $f$ & $(m$ $s^{-1})^{-1}$ $m^{-3}$ \\ \hline
        Particle source & $S$ & $(m$ $s^{-1})^{-1}$ $m^{-3}$ $s^{-1}$ \\ \hline
        Particle density & $\rho$, $\rho_g$, $\rho_r$ & $m^{-3}$ \\ \hline
        
        Momentum density & $m$ & $(m$ $s^{-1})$ $m^{-3}$ \\ \hline
        Energy density & $E$ & $(m$ $s^{-1})^2$ $m^{-3}$\\ \hline
    \end{tabular}
    \caption{Units of the parameters and variables used in the realistic application case. The value of the electronvolt is approximately $e \approx 1.60 \times 10^{-19}$. The plasma particle mass for a hydrogen plasma is approximately $m_p \approx 1.67 \times 10^{-27}$.}
 \label{table:units}
\end{table}

\begin{figure}[t!]
\centering
  \includegraphics[width=\linewidth]{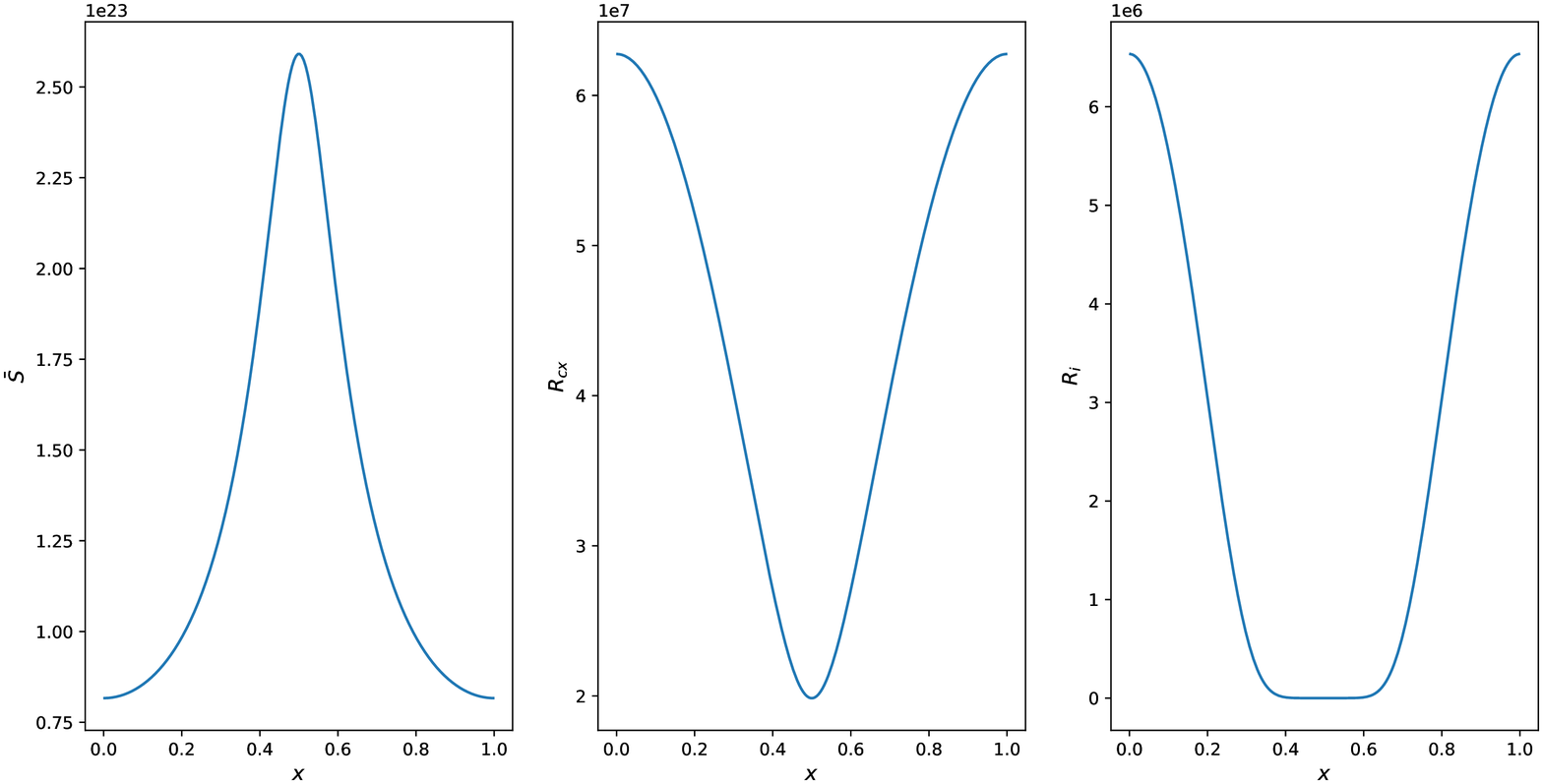}
\caption{Total source strength $\bar{S}(x)$ and collision rates $R_{cx}(x)$, $R_i(x)$ as a function of $x$ for the fictitious gas puff experiment.}
\label{fig:Exp3_Background}

  \includegraphics[width=\linewidth]{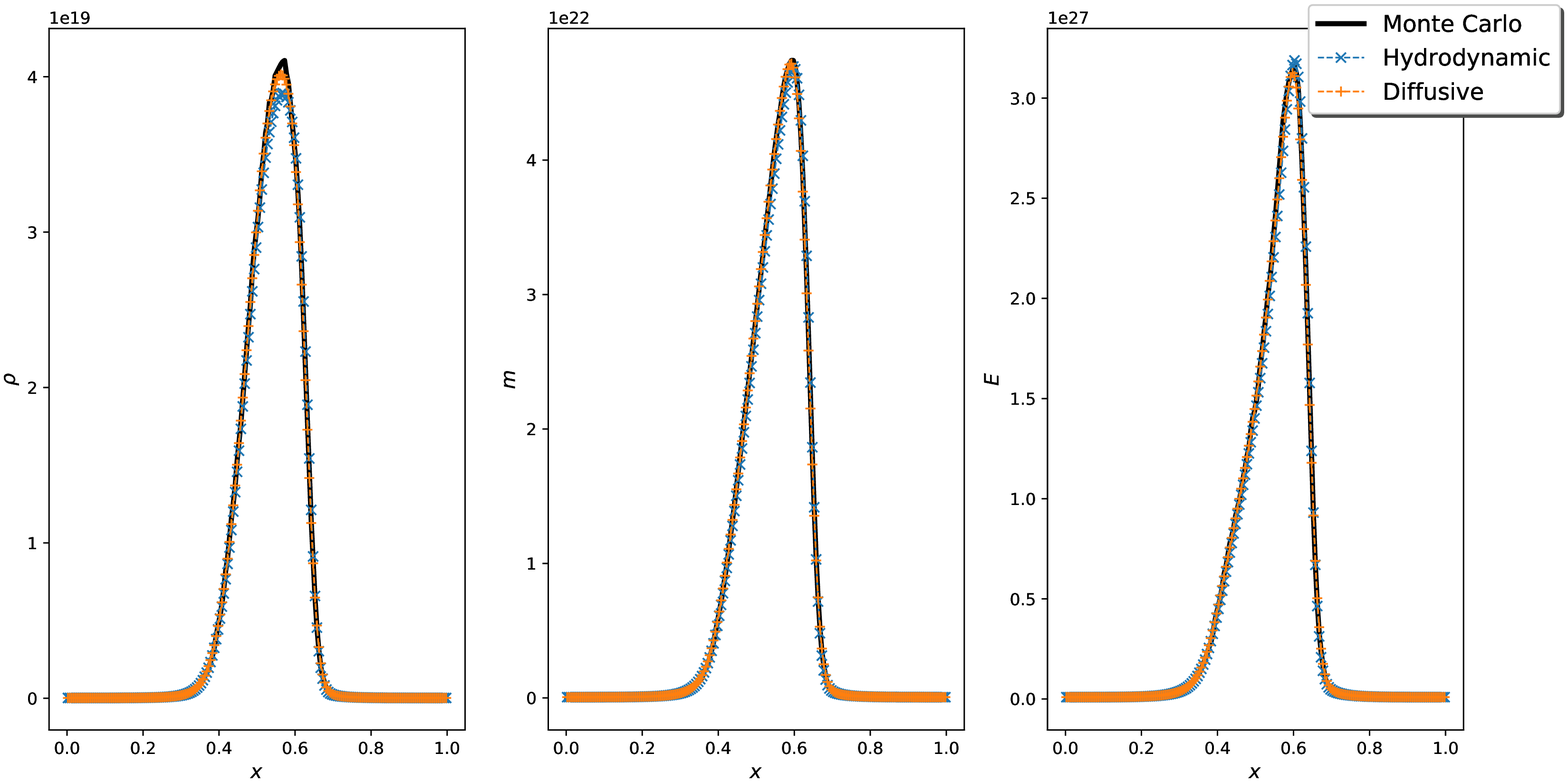}
\caption{Averaged particle, momentum and energy density of the fictitious gas puff experiment calculated using particle tracing Monte Carlo and the two Hilbert expansion based fluid models.}
\label{fig:Exp3_QoIs}
\end{figure}

The initial condition for the neutral particle density is set to zero: $\rho(x,t=0) \equiv 0$ and a stationary source $S(x,v)$ is provided. The source has two contributions: (i) a fictitious gas puff that injects instantaneously relaxing neutral particles in the plasma edge; (ii) the plasma recombination source. This results in the following expression for the source:
\begin{equation}
S(x,v) = \frac{\rho_g(x) + \rho_r(x)}{\sqrt{2 \pi \sigma_p^2(x)}} \exp\left(-\frac{1}{2}\frac{(v-u_p(x))^2}{\sigma_p^2(x)} \right),
\end{equation}
where $\rho_g(x)$ and $\rho_r(x)$ determine the spatial distribution of the gas puff source and the recombination source, respectively. For the gas puff, the spatial distribution is chosen as follows:
\begin{equation}
\rho_g(x) = 10^{21} \times \exp \left(- \frac{1}{2} \frac{(x-0.5)^2}{ 0.1^2} \right).
\end{equation}
The spatial distribution of the recombination source $\rho_r(x)$, as well as the collision rates $R_{cx}(x)$, $R_i(x)$, follow from relations taken from~\cite{horsten_comparison_2016}. These relations require the electron temperature, ion temperature, and ion density as inputs. For the temperatures, we take a profile which reaches $10$eV at the domain boundaries and equals $1$eV at the center of the domain:
\begin{equation}
T_i(x) = T_e(x) = 5.5 + 4.5 \times \cos(2 \pi x).
\end{equation}
The ion density is taken constant: $\rho_i(x) = 10^{21}$. The mean plasma velocity equals $10\%$ of the sound speed: $u_p(x) = 0.1 \times \sqrt{\sigma_p^2(x)}$, where the variance on the plasma velocity $\sigma_p^2(x)$ follows from:
\begin{equation}
\sigma_p^2(x) = \frac{eT_i(x)}{m_p}.
\end{equation} 
The total source strength $\bar{S}(x) = \rho_g(x) + \rho_r(x)$ and the collision rates $R_i(x)$, $R_{cx}(x)$ are plotted in Figure~\ref{fig:Exp3_Background}. 

We perform a simulation using the transient fluid models derived in this paper inserting the two Hilbert expansion based fluid models and the phenomenological pressure-diffusion fluid model of Refs.~\cite{horsten_fluid_2019, horsten_comparison_2016} for the conservation equation, and the particle tracing Monte Carlo reference solver using $P=10^5$ particles. The estimation time window starts at time $t_1 = 0.001$ and has a width of 1000 fluid solver time steps: $\Delta_W = 1000 \Delta t$. Because of the high variance on the plasma velocity ($\sigma^2_p \sim \mathcal{O}(10^8)$), this test case is closer to the diffusive scaling (with $\varepsilon^2 \approx 10^{-8}$) than to the hydrodynamic scaling, so the diffusive CFL number~\eqref{eq:CFL_DIFF} is used to determine the time step used by the transient fluid models. 

The simulation results are shown in Figure~\ref{fig:Exp3_QoIs}. The solution of the transient fluid models using the diffusive scaling Hilbert expansion based fluid model and the phenomenological pressure-diffusion model are indistinguishable by eye (the maximum relative difference between their energy densities is smaller than $10^{-3}$). Therefore, only the lines for the diffusive scaling Hilbert expansion based fluid model are plotted. The QoIs are peaked around the center of the domain, because there the ionization sink is the weakest and the source the strongest. The figure shows that the Hilbert expansion based fluid models, together with the employed splitting to treat the source and sink, are able to capture the QoIs in a satisfactory way. The diffusive scaling fluid model slightly outperforms the hydrodynamic scaling one, because the parameters are closer to the diffusive scaling assumptions. At first, it might seem surprising that the hydrodynamic scaling fluid model performs well, given that the test case is close to a diffusive scaling. The main cause of its accuracy is the relative importance of the source and ionization sink with respect to the conservative processes in this test case.

\section{Conclusion}
\label{sec: conclusion}
In this work, two Hilbert expansion based fluid models are derived for describing neutral particle behaviour in the bulk of the plasma edge in high-collisional regimes. The first step in the derivation of these fluid models is splitting the governing kinetic equation in a source, conservation, and sink equation. Taking moments of the source and sink equations is straightforward and leads to closed fluid equations. Taking moments of the conservation equation, however, would lead to an infinite chain of fluid equations. This problem is avoided by performing an asymptotic analysis around the equilibrium state of the kinetic equation using Hilbert expansions. The Hilbert expansions introduce physical assumptions at the kinetic description level in a clear and systematic way, elucidating the accuracy and range of validity of the resulting fluid models. Additionally, the Hilbert expansion based fluid models provide an approximation to the particle velocity distribution that is consistent with the introduced physical approximations and can be used to compute velocity dependent QoIs.

The numerical experiments show that the Hilbert expansion based fluid models perform well in high-collisional regimes when imposing periodic boundary conditions, i.e., when focusing on the bulk of the plasma edge, away from the influence of any boundary effects. A comparison with a phenomenological fluid model indicates that the Hilbert expansion based fluid model performance far from the asymptotic limit can be improved by neglecting the perturbation contribution to a quantity of interest, provided that the perturbation contribution becomes negligible in the asymptotic limit. 

In future work, the treatment of realistic boundary conditions of partial reflection type~\cite{horsten_comparison_2016, horsten_fluid_2019} will be elaborated. Ongoing research also focuses on simulation methods that are not only valid in the high-collisional regime, but over a wide range of collisionalities. The fluid models, which are accurate and computationally cheap in the high-collisional limit, are then combined with particle tracing Monte Carlo methods, which are accurate and computationally feasible in low-collisional regimes, in a hybrid fluid/Monte Carlo simulation strategy~\cite{borodin_fluid_2021, valentinuzzi_two-phases_2019, dimarco_hybrid_2008, karney_modeling_1998}.

\section*{Acknowledgement}
The resources and services used in this work were provided by the VSC (Flemish Supercomputer Center), funded by the Research Foundation - Flanders (FWO) and the Flemish Government. The authors would like to acknowledge the financial support of the CogniGron research center and the Ubbo Emmius Funds (University of Groningen). The first author is a SB PhD fellow of the Research Foundation Flanders
(FWO), funded by grant 1S64723N.
Parts of the work are supported by the Research Foundation Flanders (FWO) under project grants G078316N and G085922N.
This work has been carried out within the framework of the EUROfusion Consortium, funded by the European Union via the Euratom Research and Training Programme (Grant Agreement No 101052200 — EUROfusion). Views and opinions expressed are however those of the author(s) only and do not necessarily reflect those of the European Union or the European Commission. Neither the European Union nor the European Commission can be held responsible for them.

\appendix
\section{Pseudo-code}
\label{sec: app: pseudo-code}
The pseudo-code for the transient Hilbert expansion based fluid model simulation (Algorithm~\ref{alg:HEFluidModel}) and the transient particle tracing Monte Carlo simulation with stationary source (Algorithm~\ref{alg:transientMC_StationarySource}).
\begin{algorithm}[H]
Given the initial particle density $\rho(x_j,t=0)$ on a finite volume grid, times $t_1$ and $t_2$, and time step $\Delta t$.

\vspace{2mm}
\textbf{While $t \leq t_2$:}
\begin{itemize}
	\item[] \textbf{Simulation:}
	\begin{enumerate}
    \item Perform one time step with the source equation~\eqref{eq:source_equation} for the particle density $\rho$.
    \item Perform one time step with the Hilbert expansion based conservation equation (e.g.,~\eqref{eq:diffusive_macro_eq} for the diffusive scaling).
    \item Obtain QoIs from Hilbert expansion ansatz (e.g.~\eqref{eq:diffusive_HE_momentum} and~\eqref{eq:diffusive_HE_energy} for respectively the momentum and energy density in the diffusive scaling).
    \item Peform one time step with the sink equation~\eqref{eq:sink_equation} for all the QoIs.
	\end{enumerate}
\item[] \textbf{Estimation}: \textbf{If $t_1 \leq t \leq t_2$:}
	\begin{enumerate}
    \item[] Add contributions to time window estimators~\eqref{eq:time_window_estimators_fluid}.
	\end{enumerate}
\end{itemize}

\caption{Transient Hilbert expansion based fluid model simulation with time window estimation of QoIs.}
\label{alg:HEFluidModel}
\end{algorithm}
\vspace{1mm}
\begin{algorithm}[H]
Given the initial particle density $\rho(x_j,t=0)$ on a finite volume grid, times $t_1$ and $t_2$, and the number of particles $P$ used to construct the estimates of the QoIs.

\vspace{2mm}
\textbf{Initialization:}
\begin{enumerate}
	\item Calculate the normalization constant of the initial particle density $M_i = \int \rho(x,t=0)dx$ and of the source $M_s = t_2 \int \int S(x,v)dxdv$.
	\item Set the particle weights to $W_p = \frac{M_i + M_s}{P}$.
\end{enumerate}
\vspace{2mm}
\textbf{Iterate over the particles:}\\
\textbf{For $p = 1 \hdots P$:}
\vspace{2mm}\\
\textbf{Sample the particle:}
\begin{enumerate}
\item[] Sample with probability $\frac{M_i}{M_i+M_s}$ the particle position from $\rho(x,t=0)$ and the particle velocity from the equilibrium Maxwellian~\eqref{eq:Maxwellian}. If the particle is sampled from the initial condition, the initial time is set to zero: $t^*=0$. Sample with probability $\frac{M_s}{M_i+M_s}$ the particle position and velocity from the source $S(x,v)$. If the particle is sampled from the source, the initial time $t^*$ is sampled uniformly in $[0,t_2]$.

\end{enumerate}
\vspace{2mm}
\textbf{While $t \leq t_2$ and $W_p > 0$:}
\begin{itemize}
	\item[] \textbf{Simulation:}
	\begin{enumerate}
    \item[] Simulate the particle trajectory starting at time $t^*$ using a particle tracing Monte Carlo simulation method of choice~\cite{mortier_advanced_2020, spanier_monte_1969}.
	\end{enumerate}
\item[] \textbf{Estimation}: \textbf{If $t_1 \leq t \leq t_2$:}
	\begin{enumerate}
    \item[] Add contributions to the time window estimators~\eqref{eq:time_window_estimators_MC} using a time-integrated estimator of choice~\cite{mortier_advanced_2020, spanier_monte_1969}.
	\end{enumerate}
\end{itemize}

\caption{Transient particle tracing Monte Carlo simulation with stationary source and time window estimation of QoIs.}
\label{alg:transientMC_StationarySource}
\end{algorithm}

\textbf{Remark 5:} Note that both algorithms become more expensive for finer grids. In the fluid models, the CFL conditions~\eqref{eq:CFL_ADV},~\eqref{eq:CFL_DIFF} become more restrictive with increasing grid refinement (this is also the case for the discrete velocity model). In the particle tracing Monte Carlo method, there are two types of events~\cite{mortier_advanced_2020}: collision events, and grid cell crossing (gcc) events. The amount of collision events grows with increasing collisionality. The amount of gcc events grows with increasing grid refinement.

\section{Scaling the plasma velocity}
\label{sec: app: velocity scaling}
In the diffusive scaling, we assume that the plasma velocity $v_p \sim \mathcal{O}(1/\varepsilon)$, but the plasma velocity can take any value $-\infty < v_p < \infty$. So how do we justify this scaling? As the velocity distribution is assumed to be Maxwellian, we have that $99.994\%$ of the mass is located in the interval $[u_p \pm 4\sigma_p]$, with $u_p \sim \mathcal{O}(1)$ and $\sigma_p \sim \mathcal{O}(1/\varepsilon)$. This already shows that velocities that have a larger order of magnitude than $\mathcal{O}(1/\varepsilon)$ are highly unlikely. Furthermore, for $\varepsilon \rightarrow 0$, the region of velocities where $v_p \gg 1$ is much larger than the region of velocities where $v_p \leq 1$ because $1/\varepsilon \gg 1$, meaning that velocities of lower order of magnitude are also uncommon. It follows that both orders of magnitude lower and larger than $\mathcal{O}(1/\varepsilon)$ are unlikely, justifying our scaling $v_p \sim \mathcal{O}(1 /\varepsilon)$.

Alternatively, we can calculate the expected speed of a plasma particle $E[|v_p|]$ as follows:
\begin{equation}
\begin{split}
E[|v_p|] &= \int_{-\infty}^{\infty} \frac{|v_p|}{\sqrt{2 \pi \sigma^2_p}} \exp\left(-\frac{1}{2}\frac{(v_p-u_p)^2}{\sigma^2_p} \right) dv_p\\
&= u_p \times \erf\left(\frac{u_p}{\sqrt{2 \sigma_p^2}}\right) + \sigma_p \times \sqrt{\frac{2}{\pi}} \times \exp\left(-\frac{1}{2}\frac{u_p^2}{\sigma_p^2}\right),
\end{split}
\label{eq:expected_plasma_velocity}
\end{equation}
where we used that the plasma particles have Maxwellian~\eqref{eq:Maxwellian} as velocity distribution. Equation~\eqref{eq:expected_plasma_velocity} shows that the expected speed $E[|v_p|] \sim |u_p| + |\sigma_p|$. With $\sigma_p^2 \sim \mathcal{O}(1/\varepsilon^2)$ and $u_p \sim \mathcal{O}(1)$, it then follows that the expected speed is $|v_p| \sim \mathcal{O}(1/\varepsilon)$, again showing that the scaling $v_p \sim \mathcal{O}(1/\varepsilon)$ is justified.

\section{Exchanged momentum and energy with the plasma background}
\label{sec: app: momene exchange}
If the neutral particles were in equilibrium with the plasma background ($f(x,v,t) = f_0(x,v,t)$), then there would be no exchange of momentum and energy between the two species. The neutrals, however, are modelled as $f(x,v,t) \approx f_0(x,v,t) + \varepsilon f_1(x,v,t)$, where the perturbation $f_1(x,v,t)$ leads to an exchange in momentum and energy between the neutral particles and the plasma background. The momentum $m_{ex}$ and energy $E_{ex}$ that is being exchanged with the plasma background at time $t$ can be calculated as~\cite{mortier_advanced_2020}
\begin{equation}
\begin{split}
m_{ex}(x,t) &= R_{cx}(x,t) \int \varepsilon f_1 (x,v,t)(v-u_p(x,t))dv\\
&= R_{cx}(x,t) \int \varepsilon v f_1 (x,v,t) dv,\\
E_{ex}(x,t) &= R_{cx}(x,t) \int \varepsilon f_1(x,v,t) \left( \frac{v^2}{2} - \frac{u_p^2(x,t) + \sigma_p^2(x,t)}{2} \right) dv\\
&= R_{cx}(x,t) \int \varepsilon \frac{v^2}{2} f_1(x,v,t) dv.
\end{split}
\end{equation}
If the fluid model were exact, then the total momentum and energy, taking into account the exchange with the plasma background, would be conserved. The Hilbert expansion based fluid models, however, have an error $\sim \mathcal{O}(\varepsilon^2)$ on the momentum and energy. This error is also present in the total momentum and energy, but disappears in the limit $\varepsilon \rightarrow 0$, i.e., there is asymptotic conservation.

\section{Statistical and discretization errors}
\label{sec: app: errors}
In this Appendix, we estimate the relative statistical and discretization errors on the reference solutions and the Hilbert expansion based fluid models for Experiment 1 in Section~\ref{sec: experiments}. The relative errors are defined as in~\eqref{eq:rel_error_calculation}. The parameter $a$ is defined in Section~\ref{sec:experiment1:setup}.
\subsection{Statistical error on the particle tracing Monte Carlo method}
The particle tracing Monte Carlo reference solutions are calculated using $P=10^7$ particles and have the following estimated relative statistical errors $e_i$ on the averaged particle density $\bar{\rho}$, momentum density $\bar{m}$, and energy density $\bar{E}$:
\begin{table}[h!]
\center
\begin{tabular}{r|cc}
estimated $e_i$ & Diffusive scaling & Hydrodynamic scaling\\
	\hline
	$e_\rho$ & 0.0006 & 0.0011 \\
	$e_m$ & 0.0025 & 0.0010 \\
	$e_E$ & 0.0005 & 0.0010\\
\end{tabular}
\caption{Estimated relative statistical errors of the particle tracing Monte Carlo solution for the diffusive scaling and hydrodynamic scaling test case in Experiment 1.}
\label{table:estimatedRelativeErrors}
\end{table}

The numbers reported in Table~\ref{table:estimatedRelativeErrors} are the empirical relative standard deviations of 10 realizations of the particle tracing Monte Carlo solution for a given value of $a$, averaged over $a=0,1,\hdots 5$. The relative statistical errors are quite insensitive to the choice of $a$ (they vary less than one order of magnitude for $a=0,1,\hdots 5$). Therefore, these averaged empirical standard deviations provide a good indication of the magnitude of the statistical errors. 

\subsection{Discretization error on the Hilbert expansion based fluid models}
The relative discretization errors of the Hilbert expansion based fluid models are estimated by comparing the solution on the grid with 200 grid cells to the solution on a grid with 600 grid cells, which serves as a reference solution. For the diffusive scaling fluid model in the diffusive scaling test case in Experiment 1, the estimated relative discretization errors are given in Table~\ref{table:estimatedRelativeErrors_DiffFluid}.
\begin{table}[h!]
\center
\begin{tabular}{r|cccccc}
estimated $e_i$ & $a=0$ & $a=1$ & $a=2$ & $a=3$ & $a=4$ & $a=5$\\
\hline
$e_\rho$ & 0.0001 & 0.0001 & 0.0001 & 0.0001 & 0.0001 & 0.0001\\
$e_m$ & 0.0016 & 0.0016 & 0.0016 & 0.0016 & 0.0016 & 0.0016 \\
$e_E$ & 0.0430 & 0.0035 & 0.0004 & 0.0001 & 0.0001 & 0.0001\\
\end{tabular}
\caption{Estimated relative discretization errors of the diffusive scaling fluid model solution for the diffusive scaling test case in Experiment 1.}
\label{table:estimatedRelativeErrors_DiffFluid}
\end{table}

The estimated relative discretization errors of the hydrodynamic scaling fluid model in the hydrodynamic scaling test case in Experiment 1 are given in Table~\ref{table:estimatedRelativeErrors_HydroFluid}.
\begin{table}[h!]
\center
\begin{tabular}{r|cccccc}
estimated $e_i$ & $a=0$ & $a=1$ & $a=2$ & $a=3$ & $a=4$ & $a=5$\\
\hline
$e_\rho$ & 0.0001 & 0.0001 & 0.0001 & 0.0001 & 0.0001 & 0.0001\\
$e_m$ & 0.2078 & 0.0070 & 0.0008 & 0.0002 & 0.0001 & 0.0001 \\
$e_E$ & 0.1634 & 0.0099 & 0.0011 & 0.0002 & 0.0001 & 0.0001\\
\end{tabular}
\caption{Estimated relative discretization errors of the hydrodynamic scaling fluid model solution for the hydrodynamic scaling test case in Experiment 1.}
\label{table:estimatedRelativeErrors_HydroFluid}
\end{table}

Note that the discretization errors on the averaged density $\bar{\rho}$ are independent of $a$, because the evolution equations~\eqref{eq:diffusive_macro_eq} and~\eqref{eq:Hydrodynamical_HE_macro_eq} are independent of $\varepsilon$. The momentum density~\eqref{eq:diffusive_HE_momentum} of the diffusive scaling Hilbert expansion based fluid model is also independent of $\varepsilon$, resulting in an $a$-independent discretization error. In general, however, the discretization errors on velocity dependent QoIs do depend on $a$.

\subsection{Discretization error on the discrete velocity model}
The relative discretization errors of the discrete velocity model are estimated by comparing the solution on the grid with 200 grid cells and 200 Gauss-Hermite points to a reference solution on a grid with 600 grid cells and 600 Gauss-Hermite points. The discretization error (numerical diffusion) grows with $\sigma_p^2(x)$ in the diffusive scaling, because the Maxwellian~\eqref{eq:Maxwellian} becomes wider for larger values of $a$. We therefore only use the discrete velocity model as a reference solution in the hydrodynamic scaling test case. The estimated relative discretization errors in the hydrodynamic scaling test case in Experiment 1 are given in Table~\ref{table:estimatedRelativeErrors_DVM}.
\begin{table}[h!]
\center
\begin{tabular}{r|cccccc}
estimated $e_i$ & $a=0$ & $a=1$ & $a=2$ & $a=3$ & $a=4$ & $a=5$\\
\hline
$e_\rho$ & 0.0002 & 0.0002 & 0.0002 & 0.0002 & 0.0002 & 0.0002\\
$e_m$ & 0.0006 & 0.0006 & 0.0004 & 0.0002 & 0.0002 & 0.0002 \\
$e_E$ & 0.0008 & 0.0007 & 0.0005 & 0.0004 & 0.0004 & 0.0004\\
\end{tabular}
\caption{Estimated relative discretization errors of the discrete velocity model solution for the hydrodynamic scaling test case in Experiment 1.}
\label{table:estimatedRelativeErrors_DVM}
\end{table}

\subsection{Choice of reference solution}
In Experiment 1 in Section~\ref{sec: experiments}, the modelling error of the Hilbert expansion based fluid models in the hydrodynamic scaling test case quickly drops below the noise level of the particle tracing Monte Carlo method. Therefore, we use the discrete velocity model solution as reference solution to determine the fluid model accuracy in the hydrodynamic scaling experiment. Because the discrete velocity model error increases with $a$ in the diffusive scaling, we use the particle tracing Monte Carlo solution as reference solution to determine the fluid model accuracy in the diffusive scaling experiment.

\newpage
\bibliographystyle{abbrv}
\bibliography{Hilbert_expansion_based_fluid_models_20230405}

\end{document}